\documentclass[useAMS,usenatbib]{mn2e}
\usepackage{epsfig,amsmath,natbib}
\usepackage{color}

\def\be{\begin{equation}} 
\def\ee{\end{equation}} 
\def\ba{\begin{eqnarray}} 
\def\ea{\end{eqnarray}}

\def\cc{\,{\rm {cm^{-3}}}} 
\def\msun{{\Msun}}

\def\HH{${\rm {H_2}}\,\,$}

\def\gsim{\lower.5ex\hbox{\gtsima}} 
\def\lsim{\lower.5ex\hbox{\ltsima}} \def\gtsima{$\; \buildrel > \over 
\sim \;$} \def\ltsima{$\; \buildrel < \over \sim \;$} \def\prosima{$\; 
\buildrel \propto \over \sim \;$} \def\gsim{\lower.5ex\hbox{\gtsima}} 
\def\lsim{\lower.5ex\hbox{\ltsima}} 
\def\simgt{\lower.5ex\hbox{\gtsima}} 
\def\simlt{\lower.5ex\hbox{\ltsima}} 
\def\simpr{\lower.5ex\hbox{\prosima}} \def\la{\lsim} \def\ga{\gsim} 
  
 \def\gtsima{$\; \buildrel > \over \sim \;$} 
\def\ltsima{$\; \buildrel < \over \sim \;$} 
\def\gsim{\lower.5ex\hbox{\gtsima}} 
\def\lsim{\lower.5ex\hbox{\ltsima}} 
\def\simgt{\lower.5ex\hbox{\gtsima}} 
\def\simlt{\lower.5ex\hbox{\ltsima}} 
\def\simpr{\lower.5ex\hbox{\prosima}} 
\def\la{\lsim} 
\def\ga{\gsim}

\def\Zcr{Z_{\rm cr}}

\def\msun{\,{\rm \Msun}}

\def\E3{{\cal E}_{\rm g}^{III}}

\def\Msun{M_\odot}
\def\Zsun{Z_\odot}
\def\rvir{r_{vir}}
\def\rvir{r_{vir}}

\def\Tvir{T_{vir}} 
\def\r12{r_{1/2}} 
\def\x12{x_{1/2}} 
\def\v12{v_{1/2}}

\def\Msunyr{\, \Msun {\rm yr}^{-1}}
\title[The IMF of black holes seeds]{Initial mass function of intermediate mass black hole seeds}
\author[Ferrara et al.]{A. Ferrara$^{1,4}$, S. Salvadori$^{2}$,  B. Yue$^{1}$, D. R. G. Schleicher$^{3}$\\
$^{1}$Scuola Normale Superiore, Piazza dei Cavalieri 7, I-56126 Pisa, Italy\\
$^{2}$Kapteyn Astronomical Institute, University of Groningen, Landleven 12, 9747 AD Groningen, The Netherlands\\
$^{3}$Institut f\"ür Astrophysik, Georg-August-Universit\"at G\"ottingen, Friedrich-Hund-Platz 1, D-37077 G\"ottingen, Germany\\
$^4$Kavli Institute for the Physics and Mathematics of the Universe (WPI), Todai Institutes for Advanced Study, the University of Tokyo\\	
}

\begin{document} 
 
\date{\today} 
 
\pagerange{\pageref{firstpage}--\pageref{lastpage}} \pubyear{2012} 
 
\maketitle 
 
\label{firstpage} 
\begin{abstract} 
We study the Initial Mass Function (IMF) and hosting halo properties of Intermediate Mass Black Holes (IMBH, $10^{4-6} M_\odot$) formed inside metal-free, UV illuminated atomic cooling haloes (virial temperature $T_{vir} \ge 10^4$ K) either via the direct collapse of the gas or via an intermediate Super Massive Star (SMS) stage. These IMBHs have been recently advocated as the seeds of the supermassive black holes observed at $z\approx 6$. We achieve this goal in three steps: (a) we derive the gas accretion rate for a proto-SMS to undergo General Relativity instability and produce a direct collapse black hole (DCBH) or to enter the ZAMS and later collapse into a IMBH; (b) we use merger-tree simulations to select atomic cooling halos in which either a DCBH or SMS can form and grow, accounting for metal enrichment and major mergers that halt the growth of the proto-SMS by gas fragmentation. We derive the properties of the hosting haloes and the mass distribution of black holes at this stage, and dub it the ``Birth Mass Function''; (c) we follow the further growth of the DCBH by accreting the leftover gas in the parent halo and compute the final IMBH mass. 
We consider two extreme cases in which minihalos ($T_{vir} < 10^4$ K) can (\textit{fertile}) or cannot  (\textit{sterile}) form stars and pollute their gas leading to a different IMBH IMF. In the (fiducial) fertile case the IMF is bimodal extending over a broad range of masses, $M\approx (0.5-20)\times 10^5 \Msun$, and the DCBH accretion phase lasts from 10 to 100 Myr. If minihalos are sterile, the IMF spans the narrower mass range $M\approx (1-2.8)\times 10^6\Msun$, and the DCBH accretion phase is more extended 
($70-120$ Myr). We conclude that a good seeding prescription is to populate halos (a) of mass $7.5 < \log (M_h/\Msun) < 8$, (b) in the redshift range $8 < z < 17$, (c) with IMBH in the mass range $4.75 < (\log M_\bullet/\Msun) < 6.25$. 
\end{abstract}


\section{Introduction}
\label{Int}
Along with the formation of first stars \citep{Bromm11}, the appearance of black holes \citep{Volonteri12} is one of the most remarkable events occurring well within the first cosmic billion year (redshift $z\simgt 6$). These two types of astrophysical objects likely had a very strong impact during  cosmic evolution \citep{Ferrara08, Petri12, Park12, Jeon12, Tanaka12, Maiolino12, Valiante12} due to their radiative and mechanical energy/momentum injection in the surrounding interstellar medium of the host galaxy and into the intergalactic medium, thus drastically changing the subsequent galaxy formation history. Massive stars also disperse their metals \citep{Madau01, Johnson13, Salvadori13} in the gas, irreversibly polluting the sites of future star and black hole formation, causing a transition \citep{Bromm01, Schneider02, Omukai05, Dopke11} to the so-called PopII star formation mode we observe at lower redshifts and locally. 

For these reasons, and others discussed in the paper, the stellar and black hole populations do not evolve independently, but their formation and relative abundances are interwoven and regulated by physical processes that maintain an efficient cross-talk between them. Among these processes, suppression of \HH molecules by UV ($>0.755$ eV) and Lyman-Werner (LW, $11.2-13.6$ eV) photons is often indicated as the most important one. Molecular hydrogen is in fact a key species for gas cooling and fragmentation in the primeval Universe. On the one hand, once the intensity\footnote{Unless differently stated, we express the field intensity in the usually adopted units $J= J_{21} \times 10^{-21}$erg s$^{-1}$cm$^{-2}$Hz$^{-1}$\,sr$^{-1}$.} of the LW flux raises above a critical threshold, $J_{\nu,c}^\star$ \citep{Machacek01, Fialkov12} gas cannot cool and form stars, and consequently stellar-mass black holes, 
in minihalos that have virial temperature $\Tvir<10^4$ K. On the other hand, when larger, \textit{metal-free}, atomic-cooling ($\Tvir \simgt 10^4$ K) halos are illuminated by a sufficiently strong LW flux $J_{\nu,c} > J_{\nu,c}^\bullet$ (\citealt{Loeb94}; \citealt{Eisenstein95}; \citealt{Begelman06}; \citealt{Lodato06}; \citealt{Shang10}; \citealt{Johnson12}; \citealt{Regan09}; \citealt{Agarwal12}; \citealt{Latif13}) a ``direct collapse black hole'' (DCBH) can form\footnote{DCBH formation has however to pass through an intermediate stellar-like phase, as discussed later.  In addition, it has been suggested (e.g. \citealt{Omukai08}, \cite{Davies11}, \cite{DeVecchi12}, \cite{Miller12}) that intermediate mass black holes can also form via runaway stellar collisions in nuclear clusters. Here we do not consider such a scenario.}. The precise values of   $J_{\nu,c}^\star$ and $J_{\nu,c}^\bullet$ depend on radiative transfer, chemistry and spectral shape of the sources and they are only approximately known; however there is a broad agreement that $J_{\nu,c}^\star \ll J_{\nu,c}^\bullet = 30-1000$ in units of $10^{-21}$erg s$^{-1}$cm$^{-2}$Hz$^{-1}$sr$^{-1}$.

Recent numerical simulations and stellar evolution calculations have provided strong support in favor of the direct collapse model. In a cosmological framework, \citealt{Latif13} have shown that strong accretion flows of $ \approx 1 \Msunyr$ can occur in atomic cooling halos illuminated by strong radiation backgrounds. As stellar evolution calculations by  \citep{Hosokawa13} and  \citep{Schleicher13} suggest only weak radiative feedback for the resulting protostars, these calculations were followed for even longer time, suggesting the formation of $\simeq 10^5 \msun$ black holes \citep{Latif13b}. Their accretion can be potentially enhanced in the presence of magnetic fields, which may suppress fragmentation in the centers of these halos  \citep{Latif14}.

Even if these conditions (metal-free, atomic cooling halos illuminated by a $J_{\nu,c} > J_{\nu,c}^\bullet$ UV field) for the formation of a DCBH are met, little is known on: (i) the existence of an upper mass limit of DCBH host halos; (b) the duration of the DCBH formation/growth phase; (c) the final DCBH mass function. A fourth important question, not addressed here, concerns the final fate (e.g. inclusion in a super-massive black-hole, ejection from the host) of this intermediate ($M_\bullet = 10^{4-6} \msun$) black hole population.  These questions are at the core of a large number of cosmological and galaxy formation problems and therefore the quest for solid answers is very strong. Additional motivations come from a possible interpretation of the near-infrared cosmic background fluctuations and its recently detected cross-correlation with the X-ray background \citep{Cappelluti13}, which might imply that an unknown faint population of high-$z$ black holes could exist (\citealt{Yue13a};\citealt{Yue13}).

As we will show, answering the above questions requires a detailed description of the mass accretion and merger history of the atomic halos that satisfy the conditions described above. The process starts with the growth of a proto-SMS star inside metal-free atomic cooling halos embedded in a strong LW radiation field. The growth, fed by a high accretion rate, typically $\approx 0.1 \Msunyr$,  can be blocked by at least two type of events: the first is accretion of polluted gas, either brought by minor mergers or smooth accretion from the IGM. 

Metals would enhance the cooling rate driving thermal instabilities finally fragmenting the gas into clumps which cannot be accreted as their angular momentum is hard to dissipate. The second stopping process could be a major merger that generates vigorous turbulence, again disrupting the smooth accretion flow onto the central proto-SMS star. Note that at the rates discussed above it only takes $10^5$ yr to build a $10^4 \Msun$ SMS. If these events indeed occur, the star stops accreting and rapidly evolves toward very hot Zero Age Main Sequence (ZAMS) SMS emitting copious amonts of UV photons clearing the remaining halo gas out of the potential well. After a very brief lifetime ($< 1$ Myr) the SMS dies and leaves behind a comparable mass IMBH. If instead the star can continue to grow, it will finally encounter a General Relativity (GR) instability that will induce a rapid, direct collapse into a DCBH, i.e. without passing through a genuine stellar phase. The two cases differ dramatically, as virtually no ionizing photons are produced in the second case. Therefore the newly formed DCBH will find itself embedded in the gas reservoir of the halo and start accrete again. This accretion phase, similar to the quasi-stellar phase advocated by \cite{Begelman08}, remains highly obscured and it is only in the latest phases (several tens of Myr after the DCBH formation) that the DCBH will be able to clear the remaning gas photo-ionizing and heating it. The DCBH at that point has finally grown into a fully-fledged IMBH.  

We will investigate in detail all these steps by using a combination of analytical and numerical methods to finally derive the IMF of the IMBH. This quantity is crucial to understand the formation of supermassive black holes and the evolution of the black hole mass function with time. It also bears important implications for observations tuned to search early black hole activity. 

The question of the IMBH IMF has been tackled by some previous works. \cite{Lodato07} derive such a quantity from a stability condition for the accretion disk in a given isolated, metal-free halo by accounting for the mass that can be transported into the center without specifying in detail the physics of the BH formation. Their results are dependent on the  specific value of the critical Toomre parameter, but generally speaking they are consistent with seed masses $\simlt 10^5 \msun$. Based on the same prescriptions, \cite{Volonteri08} follow the mass assembly of SMBH resulting from such seeds up to present time using a Monte Carlo merger tree.  Perhaps closer in spirit, but fundamentally different for the physical processes involved, is the proposal by \cite{Volonteri10} that seed black holes may form via the already mentioned ``quasi-star'' phase, in which an embedded black hole forms from the collapse of a Pop III star and accretes gas at high ($\approx 1 \Msunyr$) rates. The resulting IMBH IMF peaks at a  few $\times 10^4 \Msun$, but rare supermassive seeds, with masses up to $10^6 \Msun$, are possible. This approach however, does not deal with complicating effects, included here, as the quenching of accretion due to metal pollution and feedback effects from the accreting black hole.  

Throughout the paper,  we assume a flat Universe with cosmological parameters  given by the PLANCK13 \citep{Ade13} best-fit values: $\Omega_m=0.3175$, $\Omega_{\Lambda} = 1 - \Omega_m=0.6825$, $\Omega_b h^2 = 0.022068$, and $h=0.6711$.  The  parameters defining the linear dark  matter power spectrum are $\sigma_8=0.8344$, $ n_s=0.9624$. 

\section{Physics of DCBH formation}
\label{Phy}
After the early pioneering studies \citep{Iben63, Chandrasekhar64, Loeb94, Shapiro02}, the interest in the evolution of supermassive stars (SMS) has recently received renewed attention in the context of DCBH formation. In particular, \citep{Hosokawa12, Latif13,  Johnson13a, Hosokawa13} research has focused on the previously unexplored cases of very rapid mass accretion, $\dot M = 0.1-10 \Msunyr$. 
In the following we summarize the current understanding of the aspects of SMS evolution that are relevant to the present work.

The rate at which the proto-SMS accretes gas from the surroundings plays a key role in its evolution and in particular on the stellar radius-mass relation. This can be appreciated by comparing two characteristic evolutionary timescales: the Kelvin-Helmholtz (KH),
\be
t_{KH} \equiv \frac{GM_\star^2}{R_*L_*},
\label{tkh}
\ee  
and the accretion,
\be
t_{acc} \equiv \frac{M_\star}{\dot M},
\label{tacc}
\ee  
timescales. In the early phases of the evolution, even for very small values of $\dot M$, the strong inequality $t_{KH} \gg t_{acc}$ holds, i.e. the time scale on which the star radiates its gravitational energy is much longer than the time on which mass is added to the star by accretion. Hence, the star grows almost adiabatically. As free-free absorption, providing the necessary opacity to radiative losses, depends on temperature as
$\kappa \propto \rho T^{-3.5}$, while the stellar temperature increases with mass, at some point $t_{KH} \approx t_{acc}$. After this stage, the subsequent evolution of the star depends on the accretion rate. For sufficiently low values, $\dot M \simlt 10^{-2} \Msunyr$ 
the star enters a contraction phase during which temperatures become sufficiently high to ignite hydrogen burning and the star enters
the ZAMS (for $\dot M \simlt 10^{-3} \Msunyr$ this occurs when the star has reached $M_\star \approx 50 M_\odot$) on the standard metal-free effective temperature-mass relation \citep{Bromm01a}
\be
T_{eff} = 1.1 \times 10^5 K \left(\frac{M_\star}{100 M_\odot}\right)^{0.025}.
\label{Teff}
\ee  
Note that the weak mass dependence implies that massive metal-free stars are very hot and therefore produce copious amounts of ultraviolet radiation that rapidly ionizes and clears out the remaining surrounding envelope. Thus, radiative feedback effects rapidly quench the further growth of stars once they reach the ZAMS.

The situation is drastically different for accretion rates exceeding $\dot M \simeq 10^{-2} \Msunyr$, as pointed out by \cite{Hosokawa12} and \cite{Schleicher13}: even when $t_{KH}$ has become much shorter than $t_{acc}$, the stellar radius continues to increase following very closely the mass-radius relation:
\be
R_\star = 2.6 \times 10^3 R_\odot \left(\frac{M_\star}{100 M_\odot}\right)^{1/2} \equiv R_0 \left(\frac{M_\star}{100 M_\odot}\right)^{1/2} ,
\label{Rad}
\ee  
while the effective temperature is thermostated\footnote{The thermostat is provided by the strong temperature sensitivity of H$^-$ opacity; such effect is missed if electron scattering is considered as the only source of opacity, e.g. \citet{Begelman10}.} to relatively low values, $T_{eff} \approx 5000$ K. The star then emits at a rate close to the Eddington luminosity, 
\be
L_E = \frac{4\pi G m_p c}{\sigma_T} M = L_0 \left(\frac{M}{\Msun}\right)
\ee 
where $L_0 = 1.5\times 10^{38}  {\rm \, erg\, s}^{-1}$,
as inferred from 
\be
L_\star = 4\pi R_\star^2 \sigma T_{eff}^4,
\label{Lum}
\ee  
combined with eq. \ref{Rad}.  

Why does the radius continue to grow even when the KH time has become shorter than the accretion timescale? \cite{Schleicher13} have analyzed this question in detail. The key point is that $t_{KH} \propto R_\star^{-1}$: this implies that even if at the time of formation
a shell of mass $M$ might be initially characterized by $t_{KH} \ll t_{acc}$, such inequality will be reversed as soon as its contraction begins. 
As a result, the accreting envelope phase can last considerably longer than $t_{acc}$. \cite{Schleicher13} preliminarly find that the accreting phase could continue until $M_\star=3.6\times 10^8 (\dot M/\Msunyr)\, \Msun$. Beyond this point the system will evolve into a main-sequence SMS, eventually collapsing into a black hole\footnote{Obviously the star can grow only as long as there is sufficient gas to accrete in the host halo. This point will be considered in the next Section.}. 

If this phase can be prolonged up to such high masses is a question that needs further investigation. There are hints from ongoing calculations (\citealt{Hosokawa13}; K. Omukai, private communication) that when $M_\star$ approaches $10^5 \Msun$ for accretion rates $\gsim 0.1\Msunyr$, the star enters a contraction phase (i.e. similarly to what happens for lower accretion rates at smaller masses and possibly due to a H$^-$ opacity drop); however, at this stage numerical difficulties do not allow to confirm this hypothesis. If this will turn out to be the case, the SMS final mass will be limited to $M_\star \approx 10^5 \Msun$ as the surrounding gas will be prevented from accreting by radiative feedback connected with the increased $T_{eff}$. As we will see later, the growth of proto-SMS is limited anyway by either general relativistic instabilities or cosmological effects (as for example the accretion of polluted gas, see Sec. \ref{Birth}) to masses $\simlt 3\times 10^5 \Msun$, so the above difficulties do not represent a major source of uncertainty on the final results.  

It is important to note that although D and H nuclear burning can be ignited relatively early (e.g. at 600-700 $\Msun$ for $\dot M=0.1 \Msunyr$) during the evolution, the associated energy production is however always subdominant compared to the luminosity released via Kelvin-Helmholtz contraction. Therefore it does not sensibly affect the proto-SMS evolution and final mass. The same conclusion is reached by \citet{Montero12} who performed general relativistic simulations of collapsing proto-SMS with and without rotation, including thermonuclear energy release by hydrogen and helium burning.

\begin{figure}
\includegraphics[width=90mm]{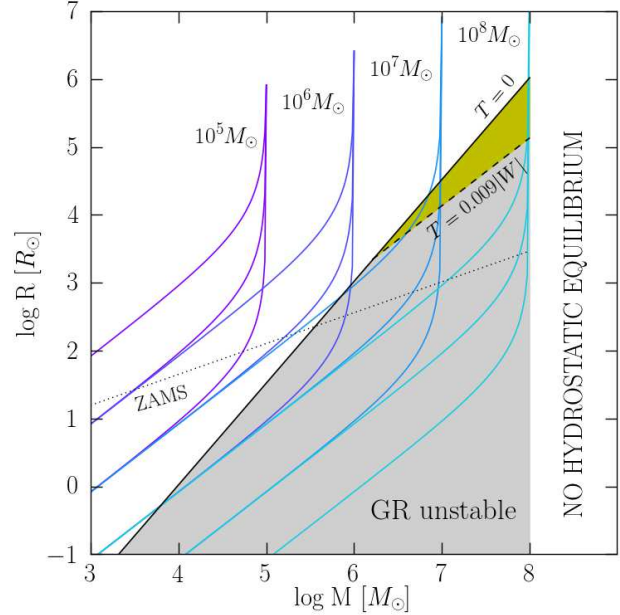}
\caption{Internal mass-radius structure of accreting supermassive protostars of different final mass $M_\star=10^{5-8} \Msun$; for each mass three different values of the accretion rate $\dot M=10, 1, 0.1 \Msunyr$ are reported from the uppermost to the lowermost set of curves. Also shown are the regions corresponding to general relativistic instability for (grey) a non-rotating ($T=0$)
and (green) for a maximally rotating ($T=0.009 \vert W\vert$) proto-SMS, along with the zero-age main sequence (ZAMS) relation. No hydrostatic equilibrium is possible beyond $10^8 \Msun$.}
\label{Fig01}
\end{figure}

The radius-mass relation for an accreting supermassive protostar can be obtained in a simple manner following the method outlined in \cite{Schleicher13}. As we have discussed above,  as long as $\dot M \simgt 10^{-2} \Msunyr$ during the accretion phase, the star radiates at a luminosity $\approx L_E$. By introducing the mass, radius and time non-dimensional variables
\be
m = M/\Msun,
\ee
\be  
r  = R/R_0, 
\ee
\be
\tau=t/t_0,
\ee  
where $t_0$ is defined starting from the KH time as
\be
t_{KH} = \frac{G\Msun^2}{R_0L_0} \frac{m}{r} \equiv t_0 \frac{m}{r} = 0.316 {\rm \, yr} \frac{m}{r},
\label{tkh0}
\ee  
a shell of accreted gas that forms at time\footnote{Note that time is simply related to mass according to $t=M/\dot M$.} $\tau_i=m/\dot m$ with initial mass $m$ and radius $r_i=m^{1/2}$ (see eq. \ref{Rad}), starts to contract according to the local KH timescale. At a generic time $t$ the stellar radius is given by
\be
\frac{dr}{d\tau} = -  \frac{r^2}{m}, 
\label{tkh1}
\ee  
or
\be
\int_{r_i}^r \frac{dr}{r^2} = -  \int_{\tau_i}^\tau\frac{dt}{m}, 
\label{tkh2}
\ee  
finally yielding the solution
\be
r(m)=\frac{m}{m^{1/2}+[(\tau(m_\star)-\tau_i(m))]}. 
\label{rm}
\ee  
The evolution or $r$ as function of $M$ for various values of $\dot M$ is shown in Fig. \ref{Fig01}. The radius grows
proportionally to $m^{1/2}$ for the inner shells, but flattens out in the envelope whose size relative to the stellar radius 
containing 90\% of the stellar mass increases with $\dot M$, becoming as large as $10^4 R_\odot$ in the most extreme case $(M_\star, \dot M)=(10^8 \Msun, 10 \Msunyr)$.  

The previous results imply that in principle the SMS final mass could be extremely large, provided a sufficiently large halo gas reservoir is present to feed it. However, the growth might be hindered by at least three factors. The first is the possible transition to the ZAMS (shown in Fig. \ref{Fig01}). In this case, as already discussed above, radiative feedback of UV photons from the now hot stellar surface will quench accretion. Although not expected, our simple treatment cannot exclude that. Detailed numerical simulations accounting for the opacity evolution will be required to address this possibility. As today, though, there is no sign of such transitions for stars that have masses up to $5\times 10^4 \Msun$. Second, the proto-SMS might instead become general relativistically (GR) unstable, directly collapsing into a black hole before reaching the ZAMS. Finally, an earlier stop to the proto-SMS growth can be imposed when the two required conditions (metal-free gas, strong UV background) for the direct collapse of the gas into a black hole cease to be valid. Whether and for how long these conditions hold can only be ascertained from a cosmological analysis that we defer to the next Section. In the following we analyze the constraints posed by the most stringent local condition, i.e. GR gravitational instability.

\subsection{Non rotating proto-SMS}
A non-rotating proto-SMS becomes gravitationally unstable (\citealt{Chandrasekhar64}, \citealt{Montero12}) when the gas equation of state (EOS) cannot be made stiff enough to compensate for the de-stabilizing general relativistic effects. Mathematically, this happens when the adiabatic index drops below the critical value
\be
\Gamma_c =  \frac{4}{3} + 1.12 \frac{R_s}{R_\star},
\label{Gam}
\ee
where $R_s = 2 GM_\star/c^2$ is the Schwarzschild radius of the star. The de-stabilizing role of the relativistic term is evident. This condition is easily translated into a critical central density\footnote{We assume a fully ionized H+He gas and adopt a He abundance $Y=0.2477$ \citep{Peimbert07}; this gives a mean molecular weight $\mu=0.59$.} at which a spherical star becomes unstable to radial perturbations:
\be
\rho_c =  1.214 \times 10^{18 }\left(\frac{\mu}{0.59}\right)^{3} \left(\frac{M_\star}{\Msun}\right)^{-7/2} {\rm g \, cm}^{-3},
\label{rhoc}
\ee
which corresponds to a mass-radius relation condition for GR instability expressed in the non-dimensional units introduced above
\be
r_c < 4.05 \times 10^{-10} m^{3/2} .
\label{unst}
\ee
The unstable region is shown as a grey area in Fig. \ref{Fig01}. As seen from there, depending on the accretion rate stars above a certain mass can become GR unstable and collapse into a black hole. By equating eq. \ref{unst} and eq. \ref{rm}  (in the limit $\tau \gg \tau_i$) we can determine the proto-SMS mass upper limit\footnote {Note that the dynamical time $(G\rho_c)^{-1/2} \ll t_{KH}$ for $M_\star \ge 10^8 \Msun$. Hence no hydrostatic equilibrium is possible beyond this mass.} 
\be
M_\star \simlt 8.48\times 10^5 \left(\frac{\dot M_\star}{\Msun {\rm yr}^{-1} }\right)^{2/3} \Msun. 
\label{unst1}
\ee
Thus, for an accretion rate of 0.15 $\Msunyr$ (typical of atomic cooling halos) a proto-SMS will collapse into a DCBH when its mass reaches $\approx 2.4 \times 10^5 \Msun$. Interestingly, this mass limit is comparable to the mass possibly marking the transition to ZAMS according to ongoing stellar evolution 1D numerical calculations. 

\subsection{Rotating proto-SMS}
If the star is rotating, this has a stabilizing effect and can hold up the collapse. In this case the expression for the adiabatic index must be modified \citep{Janka02} as follows:
\be
\Gamma_c =  \frac{2(2-5\eta)}{3(1-2\eta)} + 1.12 \frac{R_s}{R_\star},
\label{GamR}
\ee
which of course gives the correct limit (eq. \ref{Gam}) if the rotational to gravitational energy ratio, $\eta = T/\vert W \vert \rightarrow 0$. For a maximally and rigidly rotating $n=3$ polytrope, \citet{Baumgarte99} find that $\eta$ approaches the universal value 0.009 and that the instability criterion simply becomes:
\be
\left(\frac{R_\star}{R_s}\right)_c= 321, 
\ee
where we have defined the stellar radius as the equatorial one, $\approx 3/2$ times the polar radius. 
We thus obtain an equation analog to eq. \ref{unst} for the critical radius:
\be
r_c < 5.26 \times 10^{-7} m .
\label{unstR}
\ee
By equating eqs. \ref{unst} and \ref{unstR} we find that rotation increases the stability of stars with masses $M_\star^r > 1.7\times 10^6 \Msun$ (green region in  Fig. \ref{Fig01}), while at lower masses thermal pressure alone is sufficient to stabilize the star.

The results obtained in this Sec. can then be summarized by the following formulae giving the stability conditions\footnote{In principle these conditions should be complemented with the one expressing the ability of the SMS to maintain its extended accreting envelope during the growth. This relation has been derived by \citet{Schleicher13}:
$M_\star \simlt 3.6\times 10^8 \Msun (\dot M_\star/\Msun {\rm yr}^{-1})^3$, or $M_\star=1.2\times 10^6 \Msun$ for an accretion rate of 0.15 $\Msunyr$. However, in practice, this condition is met only when the star is already GR unstable for the accretion rates $>0.1 \Msunyr$ considered here, and therefore we will disregard it in the following.} for a proto-SMS:
\ba
M_\star \simlt 8.48\times 10^5 \Msun \left(\frac{\dot M_\star}{\Msun {\rm yr}^{-1} }\right)^{2/3}  \,\, (T=0), 
\ea
\ba
M_\star \simlt 6.01\times 10^5 \Msun \left(\frac{\dot M_\star}{\Msun {\rm yr}^{-1} }\right) \,\, (T\neq 0, M_\star > M_\star^{r}).
\label{stability}
\ea
proto-SMS that are more massive than the above limits will inevitably collapse to form a DCBH. 

Likely, the entire proto-SMS mass will be finally locked into the DCBH. This is confirmed by the results 
presented in \citet{Montero12} who performed general relativistic simulations of collapsing supermassive stars with and
without rotation and including the effects of thermonuclear energy released by hydrogen and helium burning. They find that 
at the end of their collapse simulation ($t \approx 10^5$ s) of a proto-SMS of mass $5\times 10^5 \Msun$, a black hole has already
formed and its apparent horizon contains a mass $\simgt 50$\% of the total initial mass. Analytical arguments discussed by 
\citet{Baumgarte99} and later refined by \citet{Shapiro02} reach a similar conclusion, indicating that $\approx 90$\% of the proto-SMS mass actually ends up into the DCBH, leaving a bare 10\% of matter in an outer region, possibly a circumstellar disk. 
This high collapse efficiency is essentially a result of the highly concentrated density profile of $n=3$ polytropes. Similar conclusions are reached by \citet{Reisswig13} who studied the three-dimensional general-relativistic collapse of rapidly rotating supermassive stars. In the following, therefore, we will make the assumption that $M_\bullet \approx M_\star$.

\section{Cosmological scenario}
\label{Sce}
\begin{figure}
\includegraphics[width=80mm]{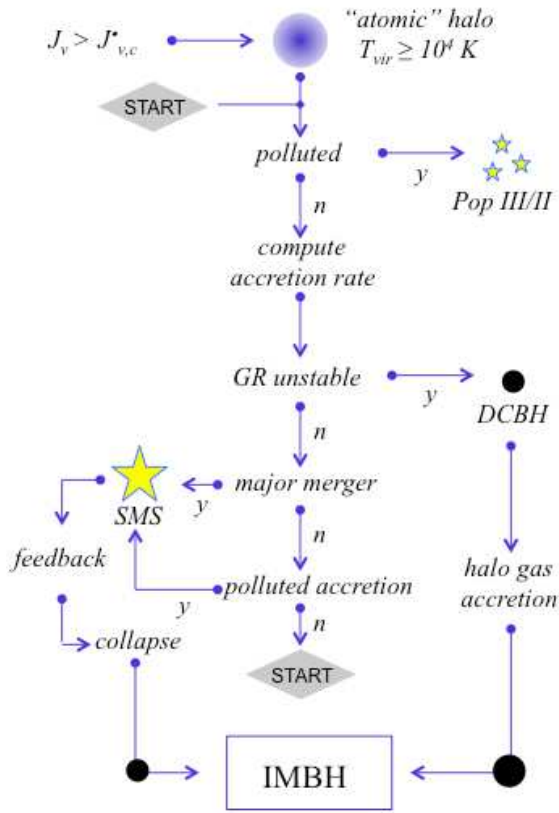}
\caption{Schematic view of the DCBH formation and growth scenario discussed in Sec. \ref{Sce}.
} 
\label{Fig02}
\end{figure}
So far we have built a physical framework for the formation of IMBH occurring either via GR instability of a proto-SMS followed by accretion, or as the end point of the evolution of a more standard SMS. We now aim at embedding such a framework in a cosmological scenario to derive the global population properties of these potential SMBH seeds. 
As the first stars form in minihalos at very high redshifts \citep{Naoz06, Salvadori09, Trenti09} their cumulative UV radiation boosted the intensity of the LW background to values $\simgt J_{\nu,c}^\star$ for which star formation in newly born minihalo is quenched. Precisely quantifying the level of suppression is difficult as it depends of several fine-grain details, although reasonable attempts have been made \citep{Ahn09, Xu13}. For these reasons, and to bracket such uncertainty, we will assume that (a) all minihalos with mass above a certain threshold corresponding to a virial temperature $T_{sf}\approx 2000$ K, or (b) none of them, form stars, i.e. radiative feedback is either moderate or extremely effective\footnote{Strictly speaking even in hypothesis (b) a small number of halos must form anyway to provide the UV radiation field.}.  We will refer to these two possibilities as the \textit{``fertile''} or \textit{``sterile''} minihalo cases, respectively.

Following the growth of cosmic structures, atomic-cooling halos start to appear as a result of accretion and merging of minihalos. In the fertile minihalo case a fraction of them are born polluted; if minihalos are instead sterile, atomic halos are metal-free by construction as no stars/metals have been produced at earlier epochs. In both cases a fraction of them will be located in regions in which $J_{\nu} > J_{\nu,c}^\bullet$ and therefore they are candidate IMBH formation sites. Recent studies  (\citealt{Ahn09}; \citealt{Yue13}) have shown that during cosmic dawn, large spatial UV field intensity fluctuations existed and persisted for long times. High illumination regions are then found near the peak of the field intensity. 

Our main goal here is to determine the Initial Mass Function of the IMBH once they are allowed to form by the environmental conditions discussed above. We will not attempt here to quantify (as done e.g. in \citealt{Dijkstra08}) how many among the unpolluted, atomic-cooling halos reside in $J_{\nu} > J_{\nu,c}^\bullet$ regions: this multiplicative (to first-order) function is only required to determine the number density of such objects. Such an assumption is equivalent to state that newly-born $\Tvir \ge 10^4$ K halos reside in highly biased regions where the field is sufficiently intense; moreover, if the first few among them manage actually to form IMBH, the radiation field of the latter will greatly amplify $J_{\nu}$, triggering the birth of additional IMBH (\citealt{Yue13}).

The formation of IMBH in atomic halos starts with an isothermal, coherent collapse centrally accumulating the gas at rates comparable or larger than the thermal accretion rate,
\be
\dot M_i \approx \frac{\pi^2}{8G} c_s^3 = 0.162 \left(\frac{T}{10^4  {\rm K} }\right)^{3/2} \Msun {\rm yr}^{-1}.
\label{tac}
\ee 
As simple as it is, this formula is in remarkable agreement with the results of most recent and complete simulations of the collapse of atomic halos. For example, by analyzing 9 such halos extracted from a cosmological large-eddy simulation, \citet {Latif13} found a very similar accretion rate of $0.1-1 \Msunyr$, measured at $z=15$, with little dependence on the galactocentric radius.

As long as this high accretion rate can be maintained, the proto-SMS continues to grow. Because of its low effective temperature, radiative feedback is unable to stop the halo gas from accreting. Eventually, the proto-SMS hits the GR unstable boundary shown in Fig. 1 and collapses into a DCBH in a very short time (about $10^5$ s). Once formed, the DCBH will continue to grow increasing its birth mass by accreting the gas leftover (if any) in their parent atomic halo finally becoming an IMBH. This feedback-regulated growth is a complex process and we will discuss it separately in Sec. \ref{Feedback}. However, if during this accretion phase the rate for any reason drops below $\approx 0.1 \Msunyr$, then the proto-SMS begins to contract and evolves into the ZAMS SMS. Given the corresponding high effective temperature (eq. \ref{Teff}), the SMS luminosity exerts a sufficient radiation pressure on the surrounding gas. Hence accretion is completely halted. As a consequence, in this case the mass of the IMBH that forms at the end of the brief ($\approx$ Myr) stellar lifetime is the same as the SMS. The main difference between these two IMBH formation channels is that for the GR instability channel, the birth mass function of DCBHs is modified during the subsequent feedback-regulated growth. This does not happen if the proto-SMS reaches the ZAMS, as already explained. For this reason we will compute the IMF of IMBH in two steps: first computing the DCBH IMF (Sec. \ref{Birth}), and then modifying it to account for the additional feedback-regulated growth.

What could cause the gas accretion rate to drop-off before the proto-SMS has become GR unstable? There are several potential show-stopper effects that could come into play. The first is that major galaxy mergers, in contrast with the smoother accretion of small lumps of matter, are likely to dramatically perturb the smooth accretion flow onto the proto-SMS. An additional effect of the merger could be that the shock-induced electron fraction enhances the cooling. This mechanism was initially proposed by \citet{Shchekinov06} and \citet{Prieto14}, and recently confirmed by a numerical simulation in the absence of radiative backgrounds \citep{Bovino14}. 

Secondly, the gas brought by the merging halos (or collected from the intergalactic medium) can be already polluted with heavy elements. As a result, clump formation following metal-cooling fragmentation of the gas is likely to drastically quench the accretion rate onto the proto-SMS, limiting its growth. 

Finally, the halo could be very gas-poor as a result of gas ejection by supernova explosions occurred in the progenitor halos. In conclusion, even if the sufficient conditions for IMBH formation in a give halo are met, the hidden and quiet growth of the proto-SMS finally leading to a DCBH via GR instability is hindered by a number of effects. All these possible physical paths are graphically summarized in Fig. \ref{Fig02}.  

The main challenge of the problem is to consistently follow the growth of a proto-SMS inside an atomic halo within a cosmological context, i.e. following the history of the parent halo as it merges with other halos and accretes gas from the intergalactic medium. We accomplish this by using a merger tree approach as described in the following.
\section{Merger trees}
\label{Mer}
In order to quantitatively investigate the above scenario we follow the merger and mass accretion history of dark matter halos 
and their baryonic component. To this aim we use the data-calibrated merger-tree code GAMETE (GAlaxy MErger Tree and Evolution, 
\citealt{Salvadori07}), which has been developed to investigate the properties of present-day ancient metal-poor stars. The code 
successfully reproduces the metallicity-luminosity relation of Milky Way (MW) dwarf galaxies, the stellar Metallicity Distribution Function (MDF) 
observed in the Galactic halo, in classical and ultra-faint dwarfs (\citealt{Salvadori09}), and the properties of very metal-poor Damped
Lyman$\alpha$ Absorbers (\citealt{Salvadori12}). Here we only summarize the main features of the code, deferring the interested 
reader to the previous papers for details. 

GAMETE reconstructs the possible merger histories of a MW-size dark matter halo via a Monte Carlo algorithm based on 
the Extended Press-Schechter theory (\citealt{Salvadori07}), tracing at the same time the star formation (SF) along the 
hierarchical trees with the following prescriptions:  (i) the SF rate is proportional to the mass of cold gas in each galaxy,
and to the SF efficiency $\epsilon_*$;  (ii) in minihalos $\epsilon_*$ is reduced as $\epsilon_{H2}=
2\epsilon_*[1+(T_{vir}/2\times 10^4$K$)^{-3}]^{-1}$ due to the ineffective cooling by H$_2$ molecules; (iii) Population II 
stars form according to a Larson IMF if the gas metallicity exceeds the critical value, $\Zcr = 10^{-5\pm1}\Zsun$
 (\citealt{Schneider06}), here assumed $\Zcr = 10^{-6} \Zsun$. At lower metallicity, PopIII stars form with reference mass 
$m_*=25\Msun$ and explosion energy $E_{SN}=10^{51}$~erg consistent with faint SNe (\citealt{Salvadori12}). 
The chemical evolution of the gas is simultaneously traced in haloes and in the surrounding MW environment by including 
the effect of SN-driven outflows, which are controlled by the SN wind efficiency. The metal filling factor, $Q_Z = V^{tot}_Z/V_{mw}$, 
is computed at each $z$ by summing the volumes of the individual metal bubbles around star-forming haloes, $V^{tot}_Z$, 
and where $V_{mw}\approx5 (1+z)^{-3}$~Mpc$^3$, is the proper MW volume at the turn-around radius (\citealt{Salvadori13}). 
The probability for newly formed halos to reside in a metal enriched region is then computed as $P(z)=[1-\exp(Q_Z)]/Q_{\delta>\delta_c}$, 
where $Q_{\delta>\delta_c}(z)$ is the volume filling factor of fluctuations with overdensity above the critical threshold, $\delta > \delta_c=1.686$, for the linear collapse (\citealt{Miralda00}). The latter quantity describes the abundance of high-density regions, in which metals first penetrate (\citealt{Tornatore07, Pallottini14}). Objects in enriched (primordial) regions are assigned an 
initial metallicity $Z_{vir}=Z_{GM}/[1-\exp(Q_Z)]$ ($Z_{vir}=0$), where $Z_{GM}$ is the average metallicity of the MW environment. 

\begin{figure*}
\includegraphics[width=85mm]{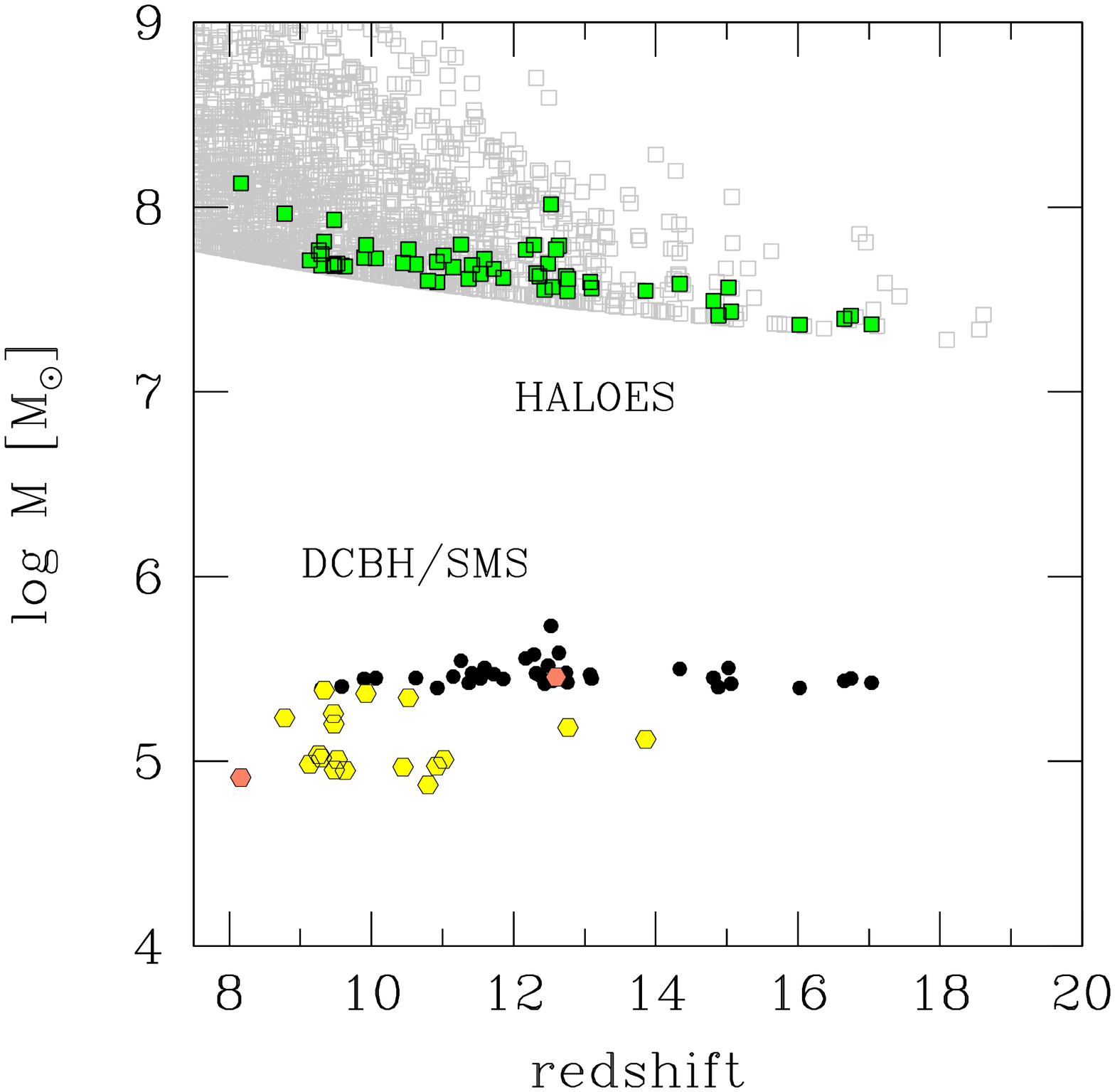}
\includegraphics[width=85mm]{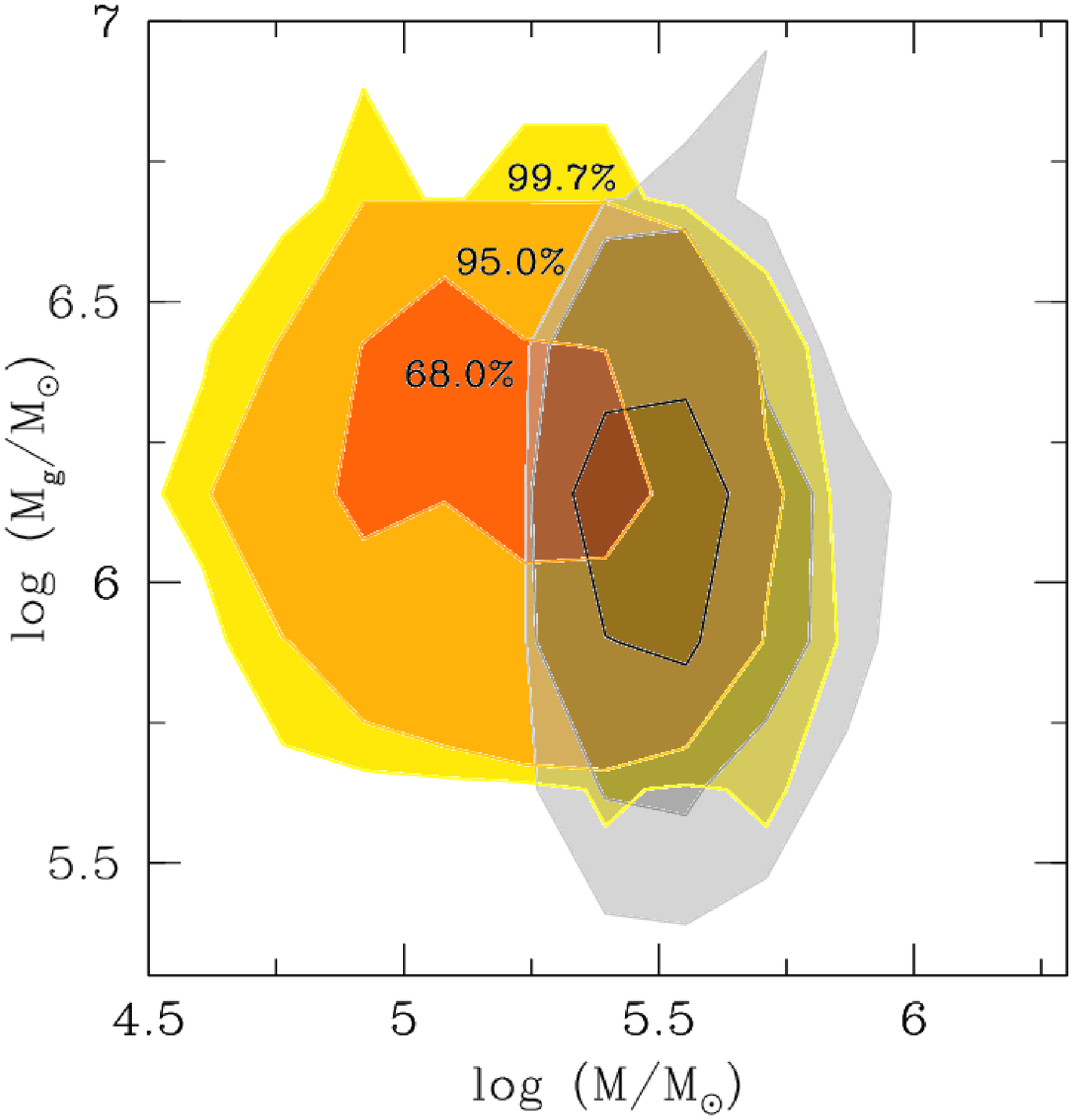}\,\,
\caption{\textbf{Left:} As a function of their formation redshift we show the mass of:
(i) halos with $\Tvir >10^4$ K (gray open squares), (ii) halos hosting a DCBH or SMS (green squares), (iii) DCBH (black circles), and (iv) SMS 
whose growth has been halted by a major merger event (red exagones) or by metal-pollution (yellow exagones). The results are shown for a single MW merger history \textbf{Right:} Mass of gas left in the halos after a DCBH (gray) or SMS (yellow-orange) formation as a function of the DCBH/SMS mass. Contours refer to 68\%, 95\%, 99.7\% confidence levels (50 realizations).   
} 
\label{Fig03}
\end{figure*}

As primordial composition ($Z<Z_{cr}$) halos in the merger tree cross the $T_{vir}=10^4$ K threshold, we postulate that a proto-SMS can form in each of them and follow its growth in the following manner. We assume that the proto-SMS is fed by an accretion rate, $\dot M_* = \max(\dot M_i,\dot M_e)$, that is the maximum between the ``internal'' accretion rate from the host halo gas, $\dot M_i$,  and the ``external'' accretion rate due to minor\footnote{For reasons explained later in this Section, we stop the proto-SMS growth after a major merger.} merger events, $\dot M_e$. $\dot M_i$ is computed as the thermal accretion rate (eq. \ref{tac}). The external rate $\dot M_e$ is taken as the ratio between the gas mass of the merging halo and the dynamical friction time-scale, $t_{merge}$, between the two colliding objects. We compute $t_{merge}$ from the classic Chandrasekhar formula \citep{Mo10}:
\be
H(z)t_{merge} = 0.234\frac{\zeta x}{\ln(1+x^2)}
\label{t_merge}
\ee
where $x=M_1/M_2$ is the ratio of the merging halo masses (with $M_1>M_2$), and $\zeta$ is the circularity parameter encoding the eccentricity of the orbit decay, which we randomly select in the interval $\zeta= [0.1,0.25]$ following \citet{Petri12}. This prescription allows the occurrence of supra-thermal accretion rates, consistently with the numerical simulation results of \citet{Mayer10} and \cite{Bonoli14} (see however \citet{Ferrara13} for a critical discussion) that show that following merger events the gas rapidly loses angular momentum and is efficiently funneled towards the nuclear region. As long as the gas in the host halo (including that brought by mergers) remains metal-free, the proto-SMS grows at a rate set by $\dot M_*$ until it eventually becomes GR unstable (eq.~\ref{unst1}).

However, GR instability is not necessarily the final fate for the proto-SMS. In fact, the protostar growth can stop because of two distinct physical processes (a) a major merger event, i.e. a collision with a halo of comparable mass, $M_1/M_2 = [0.5,2]$, and (b) pollution from heavy elements carried by a merging halo or acquired from the IGM. For reasons already explained both events are likely to suppress gas accretion onto the proto-SMS. Therefore we assume that major mergers and/or accretion of polluted gas stop accretion and lead to a zero-age main sequence SMS, which shortly after will collapse into an IMBH. To account for these events we keep track along the tree of major mergers and mergers with ``killer'' halos, i.e. halos enriched with heavy elements by previous episodes of star formation. When one of these termination events occurs, the proto-SMS growth in that halo is stopped. We store the final masses of both SMS and DCBH along with all the information on their lifetime, cosmic formation epoch and parent halo properties. Finally, we average the results over 50 random realizations of the MW-analog halo merger tree.  

\section{Birth mass function}
\label{Birth}
We are now ready to derive what we call the ``birth mass function''. This is the mass distribution including the newly formed DCBH originating 
from  GR instability of a proto-SMS, and the black holes corresponding to the end point of the SMS evolution. The final IMF of IMBH seeds (Sec. \ref{Feedback}) needs to additionally account for the subsequent feedback-regulated growth of DCBH.     

The birth mass function will be presented for the two limiting cases of fertile (representing the fiducial case) and sterile minihalos. The fertile case assumes that all minihalos with $T_{vir}>T_{sf}=2000$~K can form stars when $z\geq 10$. At lower redshifts $T_{sf}$ slowly increases up to the value $T_{sf}\approx 2\times 10^4$~K reached at $z\approx 6$, when MW environment is reionized \citep{Salvadori13}. This empirical functional form catches the essence of the increasing ability of the LW radiation to suppress star formation in halos as its intensity climbs and it is calibrated on a detailed comparison with the dwarf galaxy population of the MW halo \citep{Salvadori09}. The sterile minihalo case, instead, assumes that minihalos never form stars. These two cases are meant to bracket the uncertain role of radiative feedback in suppressing star formation via \HH destruction in these small systems. Note that in the first case atomic halos resulting from the merger of smaller progenitors can be polluted with heavy elements when they form; if instead minihalos are sterile, atomic halos are all born unpolluted.

\subsection{Fertile minihalos}
The main results for this case are depicted in Fig.~\ref{Fig03}. In the left panel we show the masses of DCBH and SMS, and of their hosting halos ($M_h$) as a function of their formation redshift for a single realization of the merger tree. For comparison purposes, the mass of all atomic halos in the merger tree at different redshifts are also shown (gray points). The halos hosting DCBH or SMS span a well-defined and narrow range of masses $M_h\approx (2-10)\times 10^7 \msun$ during the entire formation epoch, $8 < z < 17$. They are low-mass systems, which at any given redshift have roughly the minimum virial temperature required for atomic cooling, $\Tvir \approx (1-1.3) \times10^4$~K. While the chances to remain unpolluted are relatively high for these small systems (30 such halos out of a total of 36, i.e. $\approx 83\%$), this probability drops rapidly for more massive $\Tvir > 1.3\times 10^4$~K objects, which form via merging of smaller progenitors that have already formed stars. 
\begin{figure}
\includegraphics[width=80mm]{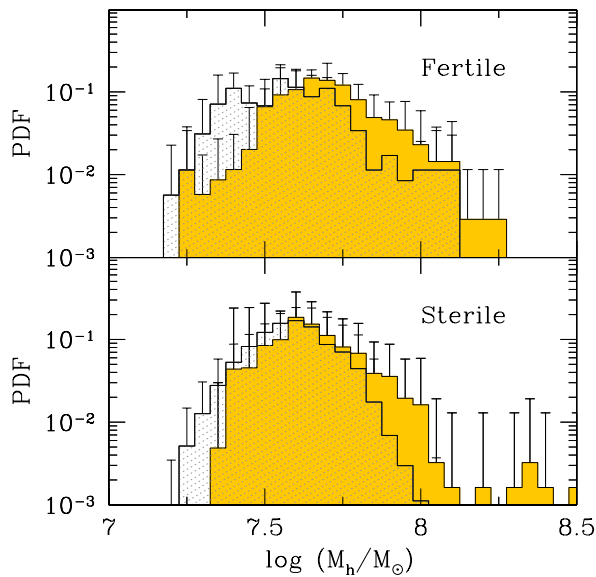}
\caption{Mass probability distribution function of DCBH (dotted histogram) or SMS (yellow shaded histogram) host halos for the fertile (upper panel) and sterile (lower panel) minihalo cases. The results are averaged over 50 MW merger histories, and the $\pm 1\sigma$ errors are shown. 
} 
\label{Fig05}
\end{figure}
These results are more quantitatively illustrated in the upper panel of Fig.~\ref{Fig05}, which shows the probability distribution function (PDF) of host halo masses. The mass distribution of DCBH hosts (gray histogram) has an almost symmetric distribution with a pronounced peak around $M_h\approx (3.5-5)\times 10^{7}\msun$ and rapidly declines towards the tails of the distribution. Most of the halos hosting DCBH ($>80\%$ of the total) are low-mass objects with $M_h \approx (2.5-6.3) \times 10^7 \msun$. Moreover, the external accretion occurs at a subdominant rate with respect to internal one, $\dot M_e = [0.0004-0.16]\leq \dot M_i$. This implies that most of the time the proto-SMS accretes gas at the thermal accretion rate set by the halo virial temperature (eq.~\ref{tac}), and hence comparable for all of them, $\dot M_* \approx (0.162-0.57)\msun$yr$^{-1}$. Since the final mass of DCBHs is entirely determined by $\dot M_*$ (see eq.~\ref{unst1}), it follows that also the DCBHs span a very narrow range of masses, $2.5\times 10^5\msun\simlt M_\bullet\simlt 4.5\times 10^5\msun$, as can be appreciated by inspecting Fig.~\ref{Fig03} (black points). 

In the same Figure we show the mass of SMS (filled exagones) whose growth has been blocked before reaching the GR instability 
because of: (i) a major merging event (red symbols), or (ii) a minor merger with a metal polluted halo (yellow symbols). It is clear from 
the Figure that metal pollution is the dominant process stopping the proto-SMS growth (22 out of 24 SMS share this origin). Moreover, 
the typical masses of SMS are smaller than DCBH, although they span a larger range, $M_{SMS}\approx (3-45)\times 10^4\msun$. This 
is due to the stochastic nature of the merging/accretion processes, which can quench the growth of the proto-SMS at different stages. 
We can also note that the formation epoch of SMS is shifted towards lower redshifts with respect to DCBH, $z\approx (14-8)$. Indeed, 
the probability to merge/accrete metal enriched gas is very low at $z\geq 15$, when only a few halos have formed stars and $Q_Z\leq 0.002$. However, it gradually increases at lower redshift becoming $\approx 1$ at $z\leq 9$, when the growth of {\it all} proto-SMS is stopped 
because of metal pollution. Due to this delay the hosting halos of SMS are typically more massive than DCBH hosts as seen in Fig.~\ref{Fig05}. 

\begin{figure*}
\includegraphics[width=85mm]{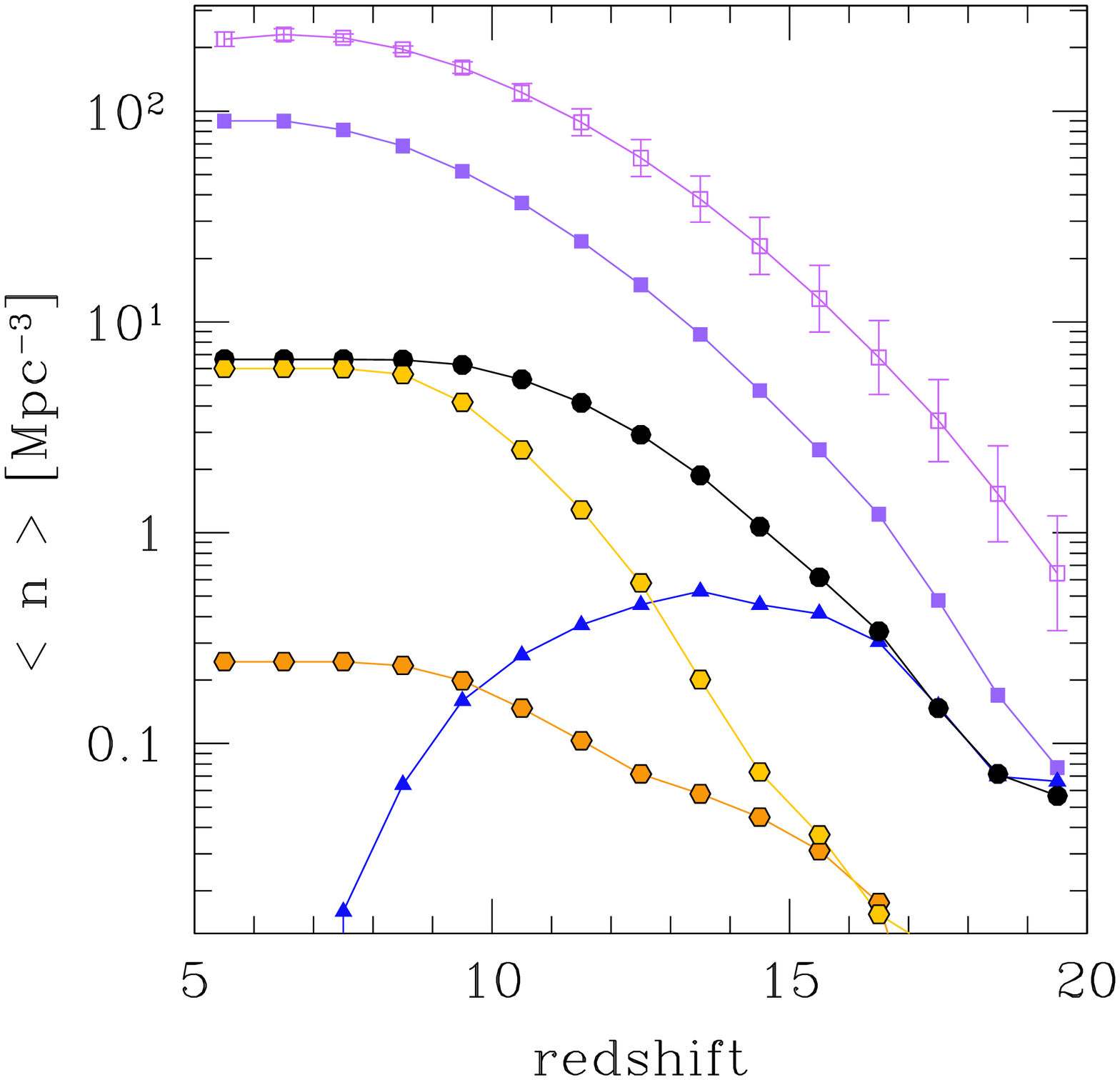}
\includegraphics[width=85mm]{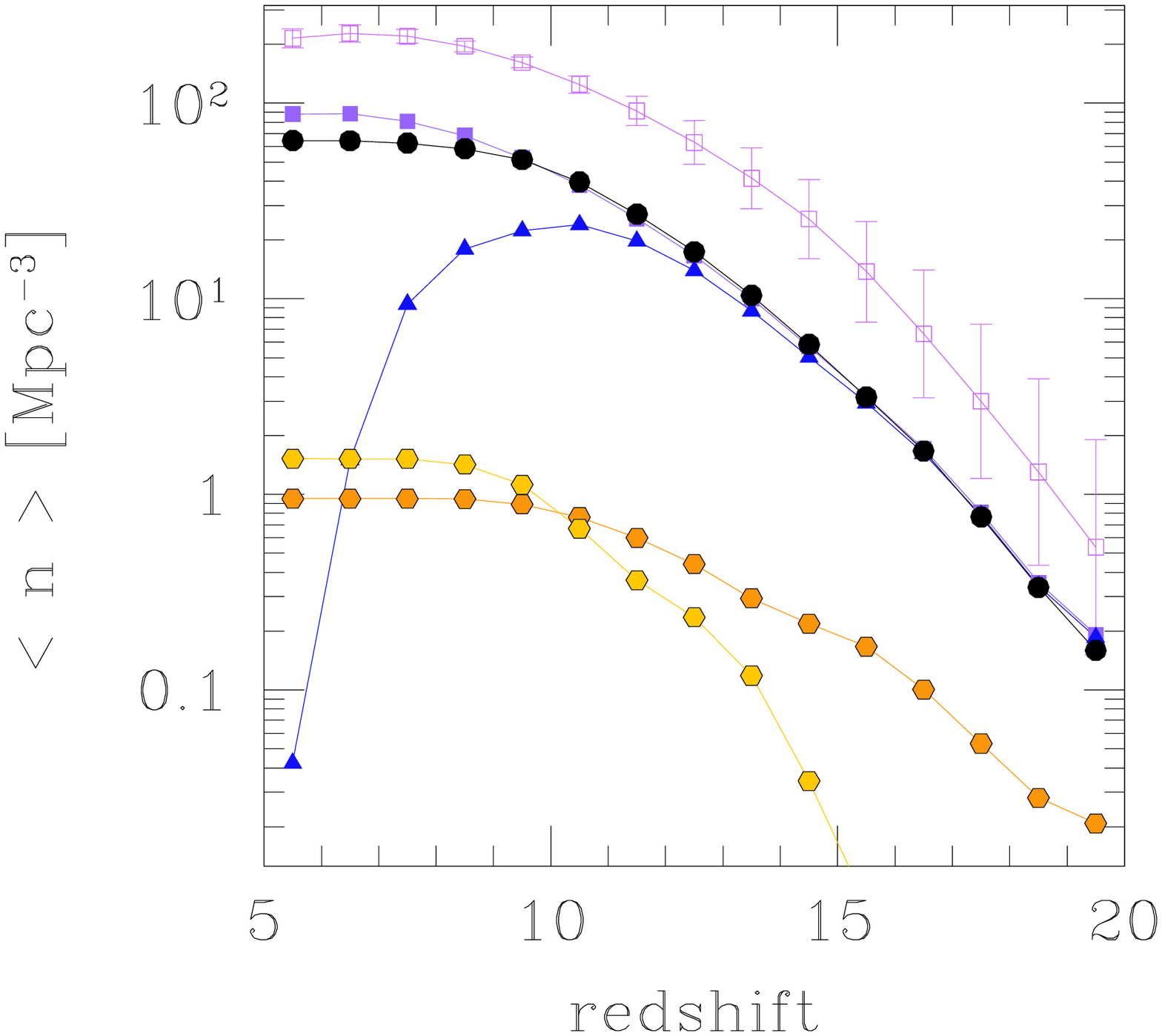}
\caption{Comoving number density evolution of: (a) all halos in the simulation (top curve), (b) halos with $\Tvir >10^4$ K (purple filled squares), (c) unpolluted ($Z< Z_{crit} = 10^{-6} \Zsun$) halos with $\Tvir >10^4$ K (blue triangles) (d) DCBH hosts (black circles), (e) SMS hosts (exagons) formed after (f) a major merger (orange) or (g) metal-pollution event (yellow). The errors are only shown for the total number of halos and represent the $\pm 1\sigma$ dispersion among different merger histories. \textbf{Left}: fertile minihalos case;  \textbf{Right}: sterile minihalos case.
} 
\label{Fig07}
\end{figure*}

All these findings can be better interpreted by inspecting Fig.~\ref{Fig07}, where the comoving number density of DCBHs (black points), 
SMS (yellow/red exagones), and halos with different physical properties (all, $\Tvir> 10^4$K, unpolluted), are shown as a function 
of redshift. It is evident that the number density, $n$, of unpolluted atomic halos (blue triangles), differently from the other curves does 
not monotonically increase with time. Instead $n$ gently grows from $z\approx 20$ to $z\approx 13$, reaches a maximum, and then 
slowly decreases when metal pollution starts to dominate. 
The number density of both DCBH and SMS are tightly connected with this evolution. From $z=20$ to $z\approx 17$ the amount 
of DCBHs increases steeply, tightly following the rise of unpolluted halos, while SMS are very rare, $n\leq 0.02$Mpc$^{-3}$. At lower 
$z$ the steepness of the curve progressively decreases, becoming flat at $z\approx 9.5$, while the SMS density progressively increases, 
gradually approaching the DCBH value. Below $z\approx 8$, also the formation of SMS is stopped, and $n$ becomes constant. At this $z$, metals have 
already reached the high density regions in which halos form, $Q_Z \approx 0.2 \geq Q_{\delta>\delta_c}$, making the onset of proto-SMS
formation impossible. The final number density of DCBHs and SMS are expected to be roughly the same, $n\approx 7$~Mpc$^{-3}$. 
We recall that this number has been obtained assuming that all host halos reside in regions in which $J_{\nu} > J^{\bullet}_{\nu,c}$, and
therefore represent a strong upper limit.

In Fig.~\ref{Fig09} we show the mass probability distribution functions of DCBH (gray histogram) and SMS (yellow histogram) normalized 
to the total number of objects (DCBH+SMS), i.e. what we call the "Birth Mass Function". The mass distribution of DCBH exhibits a peak roughly at the low-mass end, 
$M_\bullet \approx 2.5\times10^5\msun$, and monotonically declines towards higher masses. The lower limit of the PDF is populated by 
objects accreting at the thermal rate, $\dot M_* \approx 0.162$, corresponding to $T_{vir}\approx 10^4$K halos, the most common DCBH 
hosts. This sharp low-mass cut is set by GR instability (eq.~\ref{unst1}). On the other hand, more massive DCBH form in unpolluted halos 
with higher $\Tvir$, that therefore are much less common. This explaines the rapid downturn of the distribution. 

The PDF of SMS has a very different, roughly symmetric shape, displaying a wide plateau in the mass interval $M\approx (0.8-2.2)\times 10^5\msun$; it then rapidly declines towards lower/higher masses. The decreasing number of SMS with masses $< 8\times 10^4\msun$
depends on the limited number of merging/accretion (driving the proto-star evolution towards the ZAMS) occurring on timescales equal to 
$8\times 10^4\msun/\dot M_*\approx 0.1$~Myr. The PDF decline at $M>2.5\times 10^5 \msun$ is due to the same processes decribed 
for DCBH.

Another quantity that we can derive from the previous analysis is the gas left in the halo after DCBH or SMS formation. Such a quantity is 
shown as a function of DCBH/SMS mass in Fig.~\ref{Fig03} (right panel). Although DCBH masses span a very small range the gas mass 
can vary by more one order of magnitude $M_g\approx (3-50)\times 10^5\msun$. This gas can be potentially accreted by DCBH. Thus, 
as we will discuss in the next Section, the final mass of IMBH seeds will crucially depend on the subsequent accretion phase and feedback 
effects. SMS are instead able to evacuate the gas that is not quickly turned into stars. Halos residing within $68\%$ confidence level 
have $M_g=f_b (\Omega_b/\Omega_M) M_h$ with $f_b \approx 0.2$. Such a reduced gas fraction with respect to the cosmological value is the 
result of the previous star-formation activity and SN feedback processes, occurred in their progenitors. Only a few halos that form at 
$z\approx 20$, and correspond to rare high-$\sigma$ density fluctuations, are found to have $f_b\approx 1$. The halos hosting SMS 
cover roughly the same $M_g$ range as DCBH hosts. The bulk of SMS hosts, however, are more gas rich than 
DCBH hosts. This is because SMS typically form in more massive halos (see Fig.~\ref{Fig05}), which therefore contain more gas.
\subsection{Sterile minihalos}
In the second case, we consider the other extreme possibility in which the UV flux is sufficiently intense to completely suppress star formation 
in minihalos. As we commented already, strictly speaking this case corresponds to an unphysical situation as at least some stars must form in 
order to produce the required radiation field (unless some other UV source is present, as for example dark matter annihilation). The sterile 
minihalo scenario requires that the fraction of baryons converted into stars in these systems is negligible. 

With this hypothesis and caveat in mind we can analyze the results of the merger trees and the predicted properties of DCBH/SMS and of their 
hosting halos. From Fig. \ref{Fig10} we highlight two major differences with respect to the previous case. First, the formation era of DCBH/SMS stretches towards lower redshifts, $z\approx 7$. Second, the probability to merge with an already polluted halo strongly decreases. Both these effects are simply a consequence of the lack of an early metal enrichment, impying that all atomic halos are unpolluted at birth. In these objects therefore star-formation, and the subsequent metal-enrichment, is only activated if a major merger event induces a vigorous fragmentation of 
the gas, thus stopping the proto-SMS growth. 
\begin{figure}
\includegraphics[width=80mm]{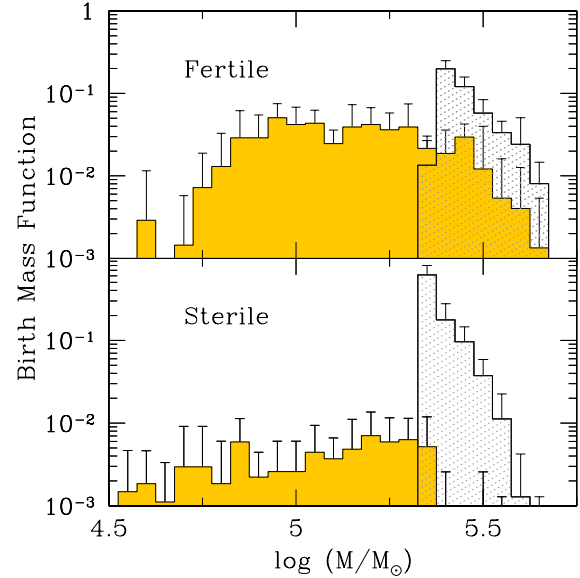}
\caption{Mass probability distribution function of DCBH seeds (dotted histogram) and SMS  (yellow shaded histogram) for the fertile (upper panel) and sterile (lower panel) minihalos case. The symbols are the same of Fig.~\ref{Fig05}.
} 
\label{Fig09}
\end{figure}

An inspection of the right panel of Fig.~\ref{Fig07} further illustrates these points. The number density of unpolluted atomic halos and 
$\Tvir > 10^4$K halos exactly overlap down to $z\approx 15$. At lower $z$, however, the two functions start to slowly deviate.  
At $z\approx 10$ the number density of unpolluted atomic halos, i.e. the sites for SMS/DCBH formation, reaches the maximumand then 
rapidly declines since metal pollution start to dominate. The maximum value is almost one order of magnitude larger than found fertile 
minihalos. As a consequence, the final number density of DCBH is much larger, $n\approx 65$~Mpc$^{-3}$. 

\begin{figure*}
\includegraphics[width=85mm]{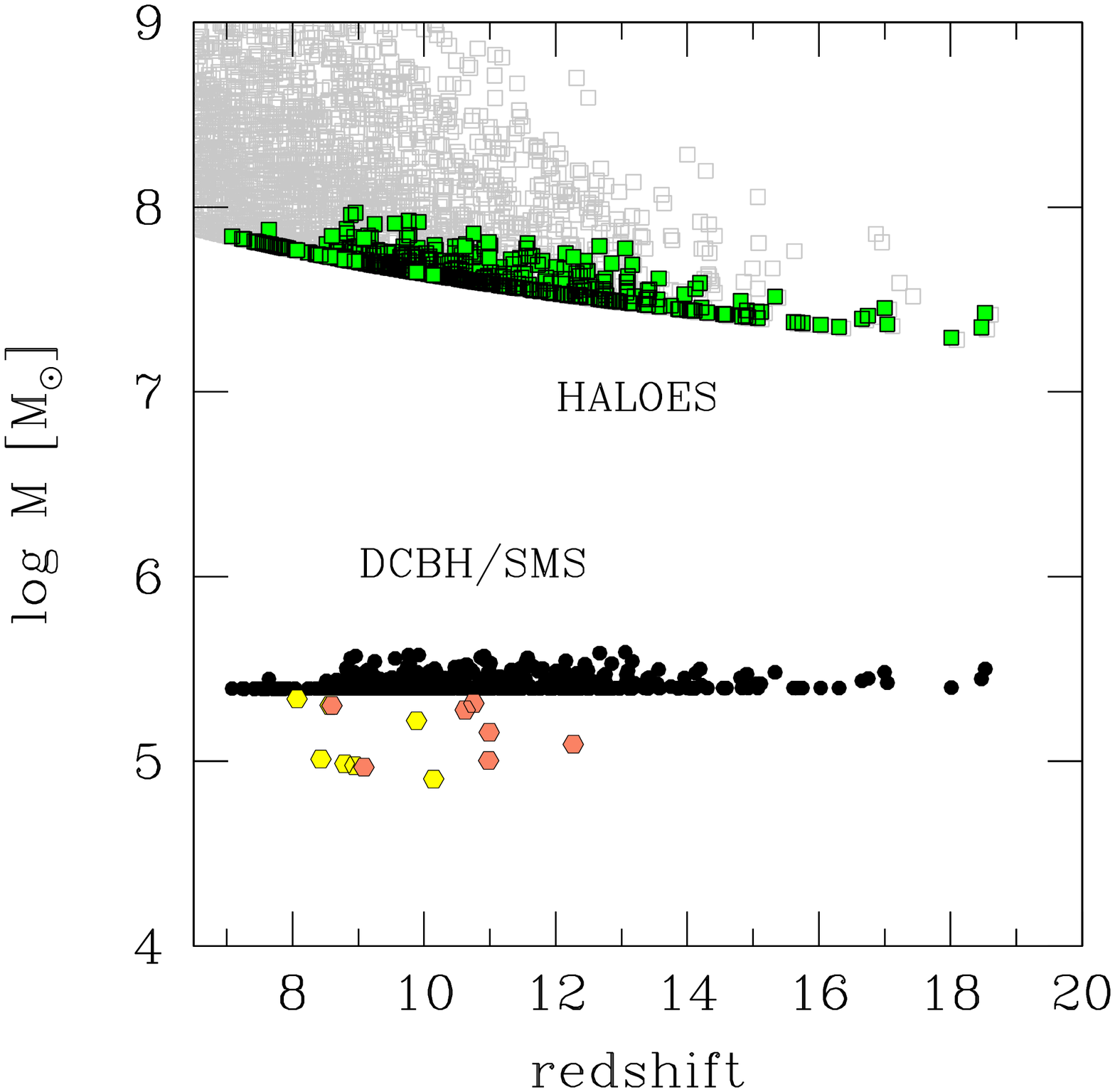}
\includegraphics[width=85mm]{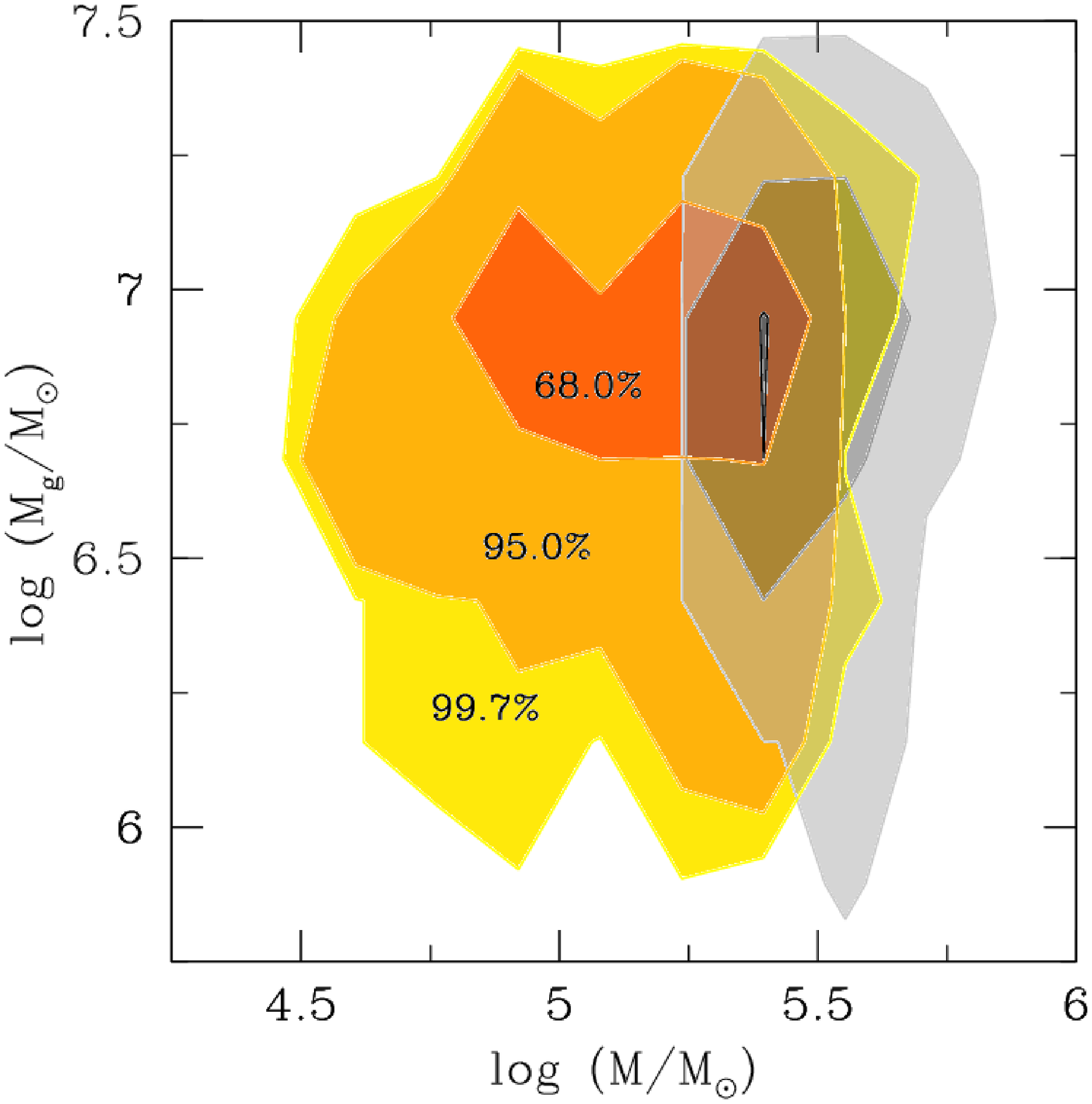}
\caption{As Fig. \ref{Fig03} for the sterile mini-halo case. 
} 
\label{Fig10}
\end{figure*}
Despite of the drastically different conditions between the fertile and sterile cases, the properties of DCBH/SMS and of their hosting halos are 
surprisingly similar. This is evident both from the mass probability distribution functions shown (Fig.\ref{Fig05}) and in the birth mass function (Fig.~\ref{Fig09}). Nevertheless there is a remarkable difference, that can be noted by comparing the right panels of Fig.~\ref{Fig10} and Fig.~\ref{Fig03}. If minihalos are sterile the gas mass at the DCBH/SMS formation is larger, $M_g\approx (1-30) \times 10^6 \msun$. At their formation, indeed, all $T_{vir}\approx 10^4$K haloes have a gas mass fraction close to $f_b\approx 1$, since no gas have been consumed within their sterile progenitor minihalos. This implies that the subsequent feedback-regulated DCBH accretion phase, will be crucial in setting the final IMF of IMBH for the two different scenarios.
\section{Feedback-regulated growth}
In order to determine the IMF of the IMBH the final step is to asses wheter they were able to accrete the gas eventually left at the time of DCBH/SMS formation. This is the goal of this Section. 
\label{Feedback}
\subsection{Formation via SMS}
If the transition to ZAMS occurs before the onset of GR instability (eq. \ref{unst}) an SMS forms. Due to the large amount of UV photons emitted by the hot ($T_{eff}\approx 10^5$ K, see eq. \ref{Teff}) stellar surface the radiation pressure on the remaining gas is very likely to evacuate it during its lifetime, expected to be $t_\star \approx 0.007 M_\star c^2/L_E = 3$ Myr, virtually independent on its mass. Obviously we cannot exclude that some fraction of this gas can be turned into stars before this happens. Irrespective of these details by the time the SMS collapses into an IMBH there will be virtually no gas left to accrete and further growth becomes impossible. In this case therefore, the IMF for these type of IMBH is the birth mass function itself.

\subsection{Formation via GR instability}
As the DCBH of mass $M_\bullet^i$ emerges from the collapse of the proto-SMS, it will be surrounded by the remaning halo gas mass, $M_g$, 
that has not been previously included into the DCBH. These two initial values are obtained from the merger tree outputs, along with the total halo mass $M_h$ (i.e. \ref{Fig03} and \ref{Fig10}). We assume that during the DCBH growth phase $M_h\approx$ const. given the short duration of such phase. In principle, all the remaining gas could be eventually incorporated into the DCBH unless feedback from energetic radiation emitted during the accretion process is able to stop or reverse the accretion flow. 

The typical density structure resulting from the isothermal collapse of the halo gas prior to DCBH formation is constituted by a central (adiabatic) core in which the collapse is stabilized, and an outer envelope where $\rho \propto r^{-2}$: 
\be
\rho(r) = \frac{\rho_c}{1+ (r/r_c)^2} .
\label{rho} 
\ee
The above density profile has been confirmed by simulations by \citet{Latif13}, who showed that the 9 candidate DCBH host halos (all of mass 
$M\approx 10^7 \Msun$) remarkably follow the distribution given by eq. \ref{rho}, independent on their mass and formation redshift (note that 
both the mass and redshift range are rather narrow as we have shown in the previous Section).  
The core radius, $r_c$, is comparable to the Jeans length of the gas $\propto c_s t_{ff}$ as in a King profile for which  
\be
r_c = \frac{3c_s}{\sqrt{4\pi G \rho_c}} = 65.5 \left(\frac{T}{10^4 \rm K}\right)^{1/2} \left(\frac{\rho_c}{10^{-11} \rm g \cc}\right)^{-1/2} \rm AU,
\label{rc} 
\ee
where the core density reference value is taken from Fig. 1 of \citet{Latif13}. The previous formula gives a core radius in very good agreement with the simulated value. Finally, we require that the mass contained within the outer radius, $r_{out}$, at which we truncate the distribution is equal to $M_g$. This gives (in the reasonable limit $r_{out} \gg r_c$),
\be
r_{out} \approx \frac{M_g^i}{4\pi\rho_c r_c^2} = \frac{GM_g^i}{9c_s^2} = 14.3\,\textrm{pc}.
\label{rout} 
\ee
Note that, due to collapse, the gas concentration increases, i.e. $r_{out}$ is more than 10 times smaller than the halo virial radius  
\be
\rvir = 583 \left(\frac{\Tvir}{10^4 \rm K}\right)^{1/2} \left(\frac{1+z}{15}\right)^{-3/2} \textrm{ pc}.
\label{rvir} 
\ee 

Assuming that, to a first approximation, the DCBH is at rest and that the accretion flow is close to spherical\footnote{The spherical approximation holds if $r_B$ is larger than the circularization radius $r_c=j^2/GM_\bullet$, where $j$ is the specific angular momentum of the gas.}, the relevant scale for accretion is the Bondi radius,
\be
r_B = \frac{2 GM_\bullet}{c_{s,\infty}^2} =  9.9 \left(\frac{M_\bullet}{10^5 \Msun}\right) \left(\frac{T}{10^4 \rm K}\right)^{-1} \textrm{ pc},
\label{rb} 
\ee 
where we denote with the subscript $\infty$ quantities evaluated at large distances from the DCBH. 
Note that $r_B$ is about 2\% of the virial radius of a typical DCBH host halo, and $r_B \approx r_{out}$, implying that the DCBH can easily drain gas from the entire volume in which gas is present. This fact has two important consequences that we analyze in the following. 

The first implication of the approximate equality between the Bondi and outer radius is that the initial gas density distribution will be
modified by the accreting DCBH. The rearrangement of the gas requires that the dynamical time is shorter than the Salpeter time, i.e.
\be
t_{ff} = \left(\frac{3\pi}{ 32 G\rho}\right)^{1/2} \ll \frac{M_\bullet}{\dot M_\bullet} \equiv t_S = 4.4\times 10^8 \epsilon \qquad \textrm{yr},  
\label{tff} 
\ee 
where we have conservatively assumed that accretion occurs at the Eddington rate. The minimum density required to re-arrange the profile fast enough is $\rho = 2.17 \times 10^{-24} \rm g\cc$, having further assumed a standard radiative efficiency $\epsilon =0.1$. As from eq. \ref{rho} we obtain that the gas density is always larger than the previous value we can safely assume that this is the case.

In order to obtain an explicit expression for the accretion flow density profile, let us proceed as follows. The one-dimensional mass and momentum conservation equations for a steady adiabatic accretion flow, i.e. the classical Bondi problem, read
\be
\frac{1}{\rho} \frac{d\rho}{dr} = -\frac{2}{r} - \frac{1}{v}\frac{dv}{dr}, 
\label{accflow1} 
\ee 
\be
v\frac{dv}{dr} + \frac{1}{\rho} \frac{dp}{dr} + \frac{GM_\bullet}{r^2} =0.
\label{accflow2} 
\ee 
Taking $c_{s,\infty}^2 = \gamma p_\infty/\rho_\infty$, where  $\gamma$ is the adiabatic index, as the sound speed at large distances, and further assuming\footnote{This may hold only approximately if the gas accretion onto the halo from the intergalactic medium is still occurring} $v_\infty = 0$, we can integrate eq. \ref{accflow2} to get the Bernoulli equation,
\be
\frac{1}{2} v^2 + \frac{c_{s}(r)^2 }{\gamma-1} -\frac{GM_\bullet}{r} = \frac{c_{s,\infty}^2 }{\gamma-1},
\label{Bernoulli} 
\ee 
from which the classical Bondi accretion rate can be derived by evaluating the previous expression at the sonic radius $r_s = GM_\bullet/2 c_s^2(r_s)$:
\be
\dot M_B = 4\pi r_s^2 \rho(r_s) c_s(r_s) = \pi q_s r_B^2 c_{s,\infty} \rho_\infty, 
\label{dotMB} 
\ee 
where 
\be
q_s(\gamma) = \frac{1}{4}\left(\frac{2}{5-3\gamma}\right)^{(5-3\gamma)/(2\gamma-2)}.
\ee 
The numerical value of $q_s$ ranges from $q_s=1/4$ at $\gamma=5/3$ to $q_s=e^{3/2}/4 \approx 1.12$ when $\gamma=1$ (isothermal); in a radiation-dominated fluid ($\gamma = 4/3$) then $q_s =\sqrt{2}/2$.

Inside the sonic radius $r_s = 1/8 r_B$ for $\gamma=4/3$, the Bernoulli equation reduces to $(1/2) v^2 \approx GM_\bullet/r$, which yields $v(r) = c_{s,\infty} (r/r_B)^{-1/2}$. To conserve the Bondi rate then the radial density dependence can be easily shown to satisfy
\be
\rho(r) = \rho_B \left(\frac{r}{r_B}\right)^{-3/2}, 
\label{rhobh} 
\ee 
where $\rho_B = 3 M_g/8 \pi r_B^3$ is a normalization constant obtained by requiring that at each time the mass contained within $r_{out}$ is equal to the current gas mass $M_g(t)$. 

Two points are worth noting. First, the $-3/2$ dependence of density is independent of the value of $\gamma$. Moreover, although it has been obtained under a steady-state assumption, it has been shown to hold also for time-dependent \citep{Sakashita74}, and even optically thick \citep{Tamazawa75} accretion flows. Thus we consider it as a robust feature of our model. In addition, as the dynamical time at small radii is much shorter than at $r_{out}$, we keep the latter fixed during the evolution and allow $\rho_B$ to decrease as gas is incorporated into the DCBH. 

From the density we can compute the optical depth to Thomson scattering (we will discuss later on when this simple opacity
prescription breaks down and adopt a more precise formulation):
\be
\tau(r) = -\int_{r_{out}}^{r}\frac{\rho(r)}{\mu m_p} \sigma_T dr =
 2 \tau_B \left(\frac{r_B}{r}\right)^{1/2}\biggr\rvert_{r_{out}}^{r}
\label{tau} 
\ee
where we have defined $\tau_B=n_B\sigma_T r_B$. Towards the center the density increases to values that are large enough to effectively trap photons; within this region the energy is convectively rather than radiatively transported by diffusion. It is easy to transform this condition (which is also equivalent to the Schwarzschild criterion for stability against convection) into one on $\tau$. The radiative luminosity can be written as
\be
L_r = - \frac{16 \pi r^2 c }{3 \kappa \rho} a T^3\frac{dT}{dr},
\label{Lr} 
\ee
where $a$ is the radiation density constant, and $\kappa=\sigma_T/\mu m_p$. On the other hand, the convective luminosity can be written as
\be
L_c = 16 \pi r^2 p v;
\label{Lc0} 
\ee
if pressure inside the trapping region is dominated by radiation, then $p = aT^4/c$. By equating the two luminosities and recalling that $d\tau=\kappa\rho dr$ we obtain the implicit definition for the trapping radius, $r_{tr}$:
\be
\tau(r_{tr})  \frac{v(r_{tr}) }{c}= 1.
\label{rtr1} 
\ee
Thus, for $r < r_{tr}$ radiation is convected inward faster than it can diffuse out and therefore within this radius (\textit{convective} region) photons cannot escape; the flow is then almost perfectly adiabatic. At larger radii, radiation can start diffuse and transport energy outwards: we refer to this region as the \textit{radiative} layer. It can be easily shown that $L_c(r_{tr}) = \beta L_E$, with $\beta= {\cal O}(1)$. Using eqs. \ref{tau} and \ref{dotMB} to express $\tau$ and recalling that $v = \dot M_{B,\gamma=4/3}/4\pi\rho r^2$, we finally obtain 
\be
r_{tr}  = \frac{\sqrt 2}{4} \tau_B \left(\frac{c_{s,\infty}}{c}\right) r_B \ll r_B,
\label{rtr} 
\ee
and 
\be
L_c = 48 \beta\left(\frac{c_{s,\infty} c}{\tau_B}\right) \dot M_B.
\label{Lc} 
\ee
The temperature at the trapping radius is then simply obtained from $T_{tr}^4 = L/\pi a c r_{tr}^2$. We will show later that the radiative region is very thin. The DCBH growth rate can then be determined by equating its accretion luminosity $\eta\dot M_\bullet c^2/(1-\eta)$ to $L_c$ to obtain 
\be
\dot M_\bullet = 48 \beta\left(\frac{1-\epsilon}{\epsilon}\right)\left(\frac{c_{s,\infty} }{c\tau_B}\right) \dot M_B.
\label{Grate} 
\ee
To determine the thermal structure of the radiative region and compute the photospheric temperature of the accreting DCBH we need to improve our treatment of the opacity. So far we have assumed a constant electron scattering opacity, $\kappa_T = \sigma_T/\mu m_p$. In a metal-free gas, this is a good approximation as long as the temperature remains $\simgt 5\times 10^4$ K. At lower temperatures, as the gas starts to recombine, additional processes increase the gas opacity: (a) free-free ($\kappa_{ff}\propto \rho T^{-7/2}$, known as the Kramers opacity); (b) bound-free and free-bound; (c) H$^-$ ($\kappa_{H^-}\propto \rho^{1/2} T^{9}$) which is mostly effective in the temperature range $(0.3-1)\times 10^4$ K. A full calculation of the opacity is given in \citet{Mayer05}; here we use a fit to their results suggested by \citet{Begelman08}:
\be
\kappa(T) = \frac{\kappa_T}{1+(T/T_*)^{-s}},
\label{Opacity} 
\ee
with $T_*=8000$ K and $s=13$. The above expression for the Rosseland mean opacity is independent on density. This turns out to be a very good approximation as long as $\rho \simlt 10^{-10} \rm g \cc$. As the use of the correct opacity becomes important outside the trapping radius where densities are comparable or below the above validity threshold, we consider this approximation as a safe and handy one. 
\begin{figure}
\includegraphics[width=80mm]{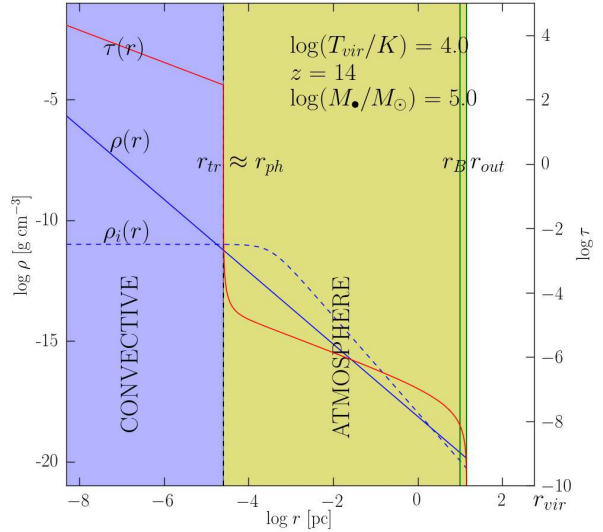}
\caption{Dependence of various characteristic scales of the problem as a function of the core gas density, $\rho_c$ for a DCBH of mass $M_\bullet=10^5 \Msun$, located in a dark matter halo with $\Tvir=10^4$ K formed at $z=14$. We have assumed $\epsilon=0.1$, $f_b=1.0$, and a radiation-dominated equation of state corresponding to $\gamma=4/3$. In addition to the core, $r_c$, Bondi, $r_B$, the outer gas distribution, $r_{out}$, and virial, $\rvir$ radii, also shown are the Bondi and Eddington density regimes corresponding to the above  ($\Tvir$, $M_\bullet$) pair and the electron scattering optical depth out to $r_B$.
} 
\label{Fig12}
\end{figure}

Armed with these prescriptions, we can solve for the temperature structure in the radiative region using the energy transport equation in the diffusion approximation:
\be
\int_{T_{tr}}^T  T'^3 \left[1+\left(\frac{T'}{T_*}\right)^{-s}\right] dT'= - \int_{r_{tr}}^r\frac{3 \kappa_T \rho(r') L }{16 \pi a c r'^2} dr',
\label{Transport} 
\ee
whose solution can be written, using the expression for the density eq. \ref{rhobh} and the definition of $\tau_B$ (eq. \ref{tau}) as 
\be
\left[ T^4 + \frac{4T_*^4}{4-s}\left(\frac{T}{T_*}\right)^{4-s}\right]_{T_{tr}}^T = \frac{3 \tau_B L}{10 \pi a c r_B^2} 
\left[\left(\frac{r_B}{r}\right)^{5/2} - \left(\frac{r_B}{r_{tr}}\right)^{5/2}\right].
\label{Solution} 
\ee
To get the photospheric radius, $r_{ph}$, we solve numerically the above equation together with the additional constrain that $\tau(r_{ph})=2/3$ as canonically used in stellar atmospheres, e.g. see \citet{Schwarzschild58}. The temperature at $r_{ph}$ is defined as the photospheric temperature of the system.

The resulting structural properties of an accreting flow onto a DCBH of mass $M_\bullet=10^5 \Msun$, located in a dark matter halo with $\Tvir=10^4$ K, $f_b=1$ formed at $z=14$ are shown, as an example, in Fig. \ref{Fig12}. We find that the radiative region is extremely narrow, 
i.e. $r_{ph} \approx r_{tr} = 2.52 \times 10^{-5}$ pc. Both radii are considerably smaller than the Bondi (9.9 pc) and virial (559.5 pc) radii. 
Inside $r_{ph}$ the optical depth raises to very large values. However, the effective temperature of the system remains relatively low, reaching in this case only 15970 K. Because of this low temperature the ionizing rate from the accreting envelope has a relatively mild feedback effect onto the overlying atmosphere, making it difficult to stop the accretion of the leftover gas onto the DCBH. 
\begin{figure}
\includegraphics[width=80mm]{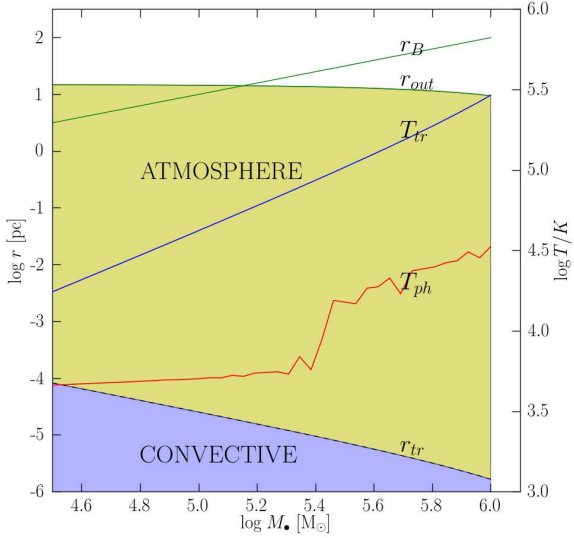}
\caption{Dependence of several characteristic radii of the system (see text for definitions) on the DCBH mass along with the temperature at the trapping radius, $T_{tr}$, (blue) and photospheric temperature, $T_{ph}$ (red). The DCBH of initial mass $M_\bullet=10^{4.5} \Msun$, is located in a dark matter halo with $\Tvir=10^4$ K formed at $z=14$. We have assumed $\epsilon=0.1$, $f_b=1.0$, and a radiation-dominated equation of state corresponding to $\gamma=4/3$.
} 
\label{Fig13}
\end{figure}

Fig. \ref{Fig13} gives a full view of the evolution of the system as the DCBH mass increases due to accretion, self-consistently calculated using eq. \ref{Grate}. As the DCBH mass increases the convective region shrinks due to the decreasing density as matter is progressively swallowed by the DCBH. At the same time, such contraction induces a temperature increase at the convective/radiative layer boundary, paralleled by a similar increase in the photospheric temperature. In particular, in this specific case of a DCBH growing inside a dark matter halo with $\Tvir=10^4$ K formed at $z=14$ and $f_b=1$, $T_{ph}$ initially increases slowly and remains below 5500 K up to the point at which the DCBH mass crosses the value $M_\bullet = 10^{5.4} M_\odot$. Beyond that point the photospheric temperature increases more rapidly and reaches about 30,000 K once the DCBH has grown to $M_\bullet = 10^{6} M_\odot$. 

Thus it is only in these more advanced evolutionary phases that copious amount of ionizing photons start to be produced. As a result
of radiative energy deposition the gas can be heated to a temperature far exceeding the virial temperature of the halo and therefore be evacuated from the halo, preventing the accretion of gas located beyond $r_{ph}$. The gas within the photosphere is eventually accreted and the final state of the system is a naked IMBH embedded in the parent dark matter halo. 
\begin{figure}
\includegraphics[width=80mm]{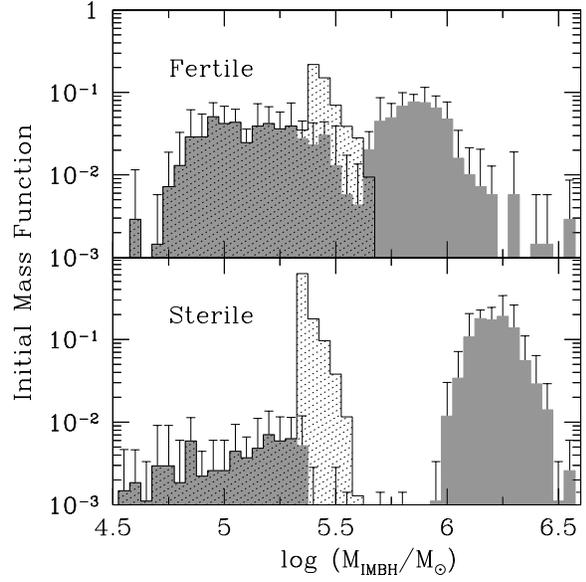}
\caption{Initial Mass Function of IMBH seeds (shaded gray histogram) averaged over 10 MW merger histories for the fertile (upper panel) and for the sterile (lower panel) minihalo cases. The errorbars correspond to $\pm 1\sigma$ errors. The birth mass function of DCBH {\it and} SMS (see Fig.~\ref{Fig09}) is also shown (dotted histogram).
} 
\label{Fig14}
\end{figure}
Thus we are left with the final question of establishing when accretion, and hence DCBH growth, will come to a halt. 

From the detailed properties computed above we derive the ionizing rate,
\be
{\cal Q}(M_\bullet)=\frac{\pi \phi(T_{ph})}{\langle h\nu\rangle} r_{ph}^2 a c T_{ph}^4
\label{Ionrate} 
\ee       
where $\langle h\nu\rangle \approx 1$ Ryd is the mean ionizing photon energy and $\phi$ is the fraction of the bolometric energy emitted by the accreting DCBH, whose spectrum is assumed to be a black-body, $B_\nu(T_{ph})$:
\be
\phi(T_{ph}) =\frac{\int_{\nu_{L}}^\infty d\nu B_\nu(T_{ph})}{\int_0^\infty d\nu B_\nu(T_{ph})} ,
\label{zeta} 
\ee       
where $h\nu_L=1$ Ryd. In order to ionize the entire atmosphere (i.e. the gas outside $r_{ph}$) and increase the gas temperature above $T_{vir}\approx 10^4$~K, the ionization rate must exceed the recombination rate, ${\cal R}$, of the gas within $r_{out}$. The latter can be written as
\be
{\cal R}(M_\bullet)=4\pi \int_{r_{ph}}^{r_{out}} dr \left(\frac{r^2}{t_{rec}} \right) \left(\frac{\rho}{\mu m_p}\right) 
\label{Recrate} 
\ee       
where $t_{rec}=(n\alpha^{(2)})^{-1}$ is the recombination timescale and $\alpha^{(2)} = 2.6\times 10^{-13} (T/10^4{\rm K})^{-1/2}$ is the Case B recombination rate of hydrogen \citep{Maselli03}.  By substituting eq. \ref{rhobh} and performing simple algebra we obtain the final expression for the recombination rate:
\be
{\cal R}(M_\bullet)=\frac{4 \pi \alpha^{(2)} \rho_B^2}{(\mu m_p)^2} r_B^3 \ln\left(\frac{r_{out}}{r_{ph}}\right) .
\label{Recrate1} 
\ee
Once the condition ${\cal Q} > {\cal R}$ is satisfied, we assume that the remaining gas has been heated and ejected by the accreting DCBH radiative feedback and its growth is quenched. This sets the final mass of the DCBH, or the mass of the resulting IMBH. 

\begin{figure}
\includegraphics[width=80mm]{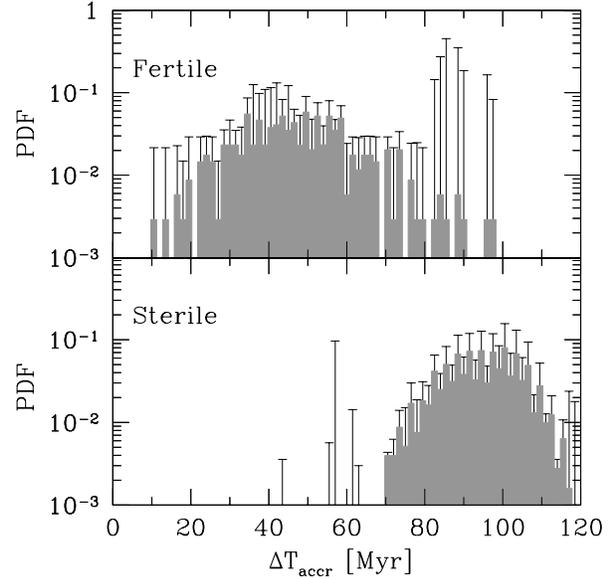}
\caption{Probability distribution functions of the accretion phase duration for the fertile (upper panel) and the sterile (lower panel) minihalo cases.
The results are averaged over 10 MW merger histories and the $\pm 1 \sigma$ errors are shown.} 
\label{Fig15}
\end{figure}
In order to determine the IMF of IMBH we follow the feedback-regulated growth of DCBH present 
in 10 realizations of the merger tree by assuming that the hosting halo mass remain constant during 
this phase (no merging processes). The final mass distribution of DCBH is then summed with the birth 
mass function of SMS (as these objects also finally evolve into black holes, see Section 5), and normalized. The results of this calculation are shown in 
Fig.~\ref{Fig14}, where the final IMF of IMBH (gray shaded histograms) is compared with the birth mass 
function of DCBH and SMS (dotted histograms). The low-mass end of the IMF is identical to the birth 
mass function, while the peak is shifted towards higher masses. This is simply a consequence of the feedback-regulated growth, only affecting DCBH. As the growth of DCBH is fed by the available halo 
gas, the displacement is larger when minihalos are sterile because in this case DCBH hosts are more 
gas rich (Fig.~\ref{Fig09} and Fig.~\ref{Fig03}, right panels). 

As a consequence, the IMBH IMF is very different in the two scenarios: in the fertile case, in particular, it exhibits a bimodal distribution with two separate peaks at $M\approx (0.7-1.2)\times 10^5\Msun$ and $M\approx (5-10)\times 10^5\Msun$. The distribution extends over a broad range of masses, from $M\approx (0.5-20)\times 10^5 \Msun$. If minihalos are sterile, the IMF  spans the narrower mass range, $M\approx (1-2.8)\times 10^6\Msun$, which contains $>90\%$ of the IMBH population. 

These differences are also reflected in the duration of the accretion phase. DCBH can continue to grow almost unimpeded for several tens of Myr before gas accretion is shut down by feedback, as illustrated by Fig. \ref{Fig15}. From there we see that in the fertile case there is a spread in the duration of the accretion phase from 10 to 100 Myr, which arises from a combination of differences in the initial DCBH mass and, more importantly, in the amount of gas available set by the past history of the host halo. The accretion phase duration distribution peaks at around 40-60 Myr. In the sterile case, durations are both longer ($70-120$ Myr, due to the larger reservoir of gas available for accretion in the halo) and more concentrated (atomic halos have similar gas content, $f_b \approx 1$, due to the lack of star formation and supernova feedback in the progenitor minihalos). 
\begin{figure}
\includegraphics[width=80mm]{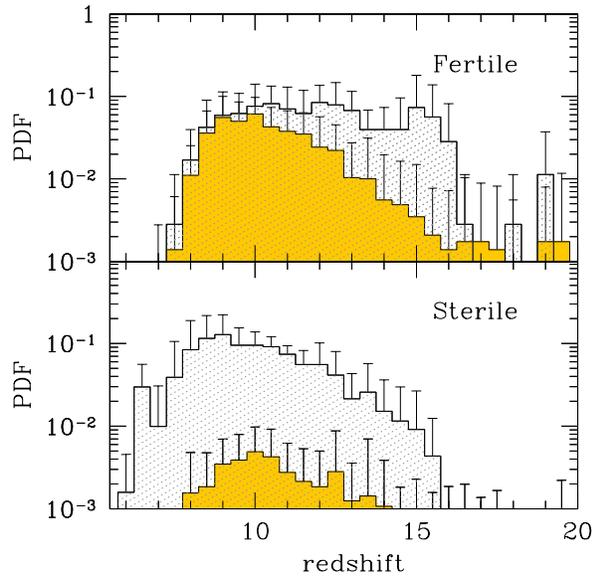}
\caption{Probability distribution function of the formation redshifts DCBH (dotted histogram) and SMS (yellow shaded histogram) for the fertile (upper panel) and sterile (lower) minihalo cases are shown. The results are averaged over 10 MW merger histories.
The errorbars correspond to $\pm 1\sigma$ errors.}
\label{Fig06}
\end{figure}
We have also derived the distribution of the formation epoch of DCBH and SMS, shown in Fig. \ref{Fig06}. Remarkably, the conditions for the formation of these black holes seeds last relatively shortly during cosmic history. In the fiducial fertile minihalos case, the first DCBH and SMS (showing a more gradual abundance rise) appear in non-negligible numbers at $z=17$; however by $z=8$ their formation is already quenched as a result of the accretion of polluted gas and/or a major merging disrupting the quiet accretion flow and inducing gas fragmentation. If minihalos are sterile, then the termination epoch is delayed by about 2 redshift units, and DCBH become the dominant source of production for IMBH seeds.

We finally comment on the relation between the IMBH and their host halo mass. This relation is often necessary to formulate physical seeding prescriptions, e.g. in studies of SMBH formation based on merger trees or numerical simulations. Our results show that a very reasonable prescription is to populate a given fraction of halos  (a) of mass $7.5 < \log M_h < 8$, (b) in the redshift range $8 < z <17$, (c) with IMBH in the mass range $4.75 < \log M_\bullet < 6.25$. This prescription assumes our fiducial case of fertile minihalos. If instead one wishes to consider sterile minihalos, then the previous prescription turns in the following set:  
(a) $7.25 < \log M_h < 7.75$, (b) in the redshift range $6 < z < 14$, (c) with IMBH in the mass range $6 < \log M_\bullet < 6.25$. We recall once again that the above fraction of such halos to be populated cannot 
be obtained from our method as it would require a detailed knowledge of the LW UV background field. Therefore such information must be fixed from other physical considerations or left as a free parameter.

\section{Summary and discussion}
\label{Sad}
In this paper we have derived for the first time the Initial Mass Function of Intermediate Mass Black Holes ($10^{4-6} M_\odot$) formed inside metal-free, UV illuminated atomic cooling (virial temperature $T_{vir} \ge 10^4$ K) halos either via direct collapse followed by GR instability or via an intermediate Super Massive Star (SMS) stage. These objects have been recently advocated as the seeds of the supermassive black holes observed at $z\approx 6$. Assembling the SMBH mass ($M_\bullet=2\times 10^9 M_\odot$) deduced for the most distant quasar ULAS J1120+0641 at $z=7.085$ \citep{Mortlock11} when $t(z) = 0.77$ Gyr, requires a seed mass $> 400 M_\odot$. Such value is uncomfortably large when compared to the most recent estimates of the mass of first stars, which now converge towards values $\ll 100 M_\odot$ \citep{Greif11, Hosokawa12a, Hirano14}. This is why IMBH seeds, with their larger masses, are now strongly preferred as the most promising seeds.

We have obtained the IMBH IMF with a three-step strategy, as described below.
\begin{itemize}
\item 
We have first derived the condition for a proto-SMS to undergo GR instability and directly collapse into a DCBH depending on the gas accretion rate; we found that, for a non-rotating SMS, GR instability kick in when the stellar mass reaches 
\be
M_\star = 8.48\times 10^5 \left(\frac{\dot M_\star}{\Msun {\rm yr}^{-1} }\right)^{2/3} \Msun. 
\ee
Thus, for an accretion rate of 0.15 $\Msunyr$ (typical of atomic cooling halos) a proto-SMS will collapse into a DCBH when its mass reaches $\approx 2.4 \times 10^5 \Msun$. A similar expression has been obtained for rotating SMS, and given by eq. \ref{stability}.

However, the SMS growth can come to an end before the star crosses the above critical mass. This occurs if the host halo accretes polluted gas, either brought by minor mergers or smooth accretion from the IGM, or suffers a  major merger that generates vigorous turbulence, again disrupting the smooth and quiet accretion flow onto the central proto-SMS star.

\item
We followed these processes in a cosmological context using the merger tree code GAMETE, which allows us to spot metal-free atomic cooling halos in which either a DCBH or SMS can form and grow, accounting for their metal enrichment and major mergers that halt the growth of the proto-SMS by gas fragmentation. We derive the mass distribution of black holes at this stage, and dub it the ``Birth Mass Function'' (BMF). Most DCBH host halos ($>80\%$ of the total) have $M_h \approx (2.5-6.3) \times 10^7 \msun$. As a result of accretion physics, DCBHs span a very narrow range of masses, $2.5\times 10^5\msun\simlt M_\bullet\simlt 4.5\times 10^5\msun$.
We find that the metal pollution is by far the dominant process stopping the proto-SMS growth. The resulting SMS are smaller than DCBH, although they span a larger range, $M_{SMS}\approx (3-45)\times 10^4\msun$, due to the stochastic nature of the merging/accretion processes. We can also note that the formation epoch of SMS is shifted towards lower redshifts with respect to DCBH, $8 < z < 14$ instead of $8 < z < 17$. The previous results refer to the fiducial (\textit{fertile}) case in which minihalos ($T_{vir} < 10^4$ K) can form stars and pollute their gas. Results are also given and discussed for the sterile case in Sec. \ref{Birth}. 

\item
As a third and final step towards the IMBH IMF we have followed the accretion of the halo gas leftover after the formation of the DCBH
onto the DCBH itself. This is necessary because, contrary to the case of the SMS in which ionizing radiation from the exposed hot photosphere ionized and disperses the surrounding gas, the General Relativity (GR) instability induces a rapid, direct collapse into a DCBH, i.e. without passing through a genuine stellar phase. The two cases differ dramatically, as virtually no ionizing photons are produced if a DCBH forms. Therefore the newly formed DCBH will find itself embedded in the gas reservoir of the halo and start accrete again. This accretion phase, similar to the quasi-stellar phase advocated by \cite{Begelman08} (see also \cite{Ball12a}), remains highly obscured and it is only in the latest phases (several tens of Myr after the DCBH formation) that the DCBH will be able to clear the remaining gas when the photospheric temperature starts to climbs from about 5000 K when DCBH mass crosses the value $M_\bullet = 10^{5.4} M_\odot$. Beyond that point the photospheric temperature increases rapidly and reaches about 30,000 K once the DCBH has grown to $M_\bullet = 10^{6} M_\odot$, thus allowing radiative feedback to clear the gas, stop accretion, and determine the final IMBH mass.
\end{itemize}

The IMBH IMF is different in the two scenarios considered: in the (fiducial) fertile case it is bimodal with two broad peaks at $M\approx (0.7-1.2)\times 10^5\Msun$ and $M\approx (5-10)\times 10^5\Msun$. The distribution extends over a wide range of masses, from $M\approx (0.5-20)\times 10^5 \Msun$ and the DCBH accretion phase lasts from 10 to 100 Myr. If minihalos are sterile, the IMF spans the narrower mass range $M\approx (1-2.8)\times 10^6\Msun$ containing $>90\%$ of the IMBH population (the remaining part being represented by the SMS low mass tail, see Fig. \ref{Fig14}). 
We conclude that a good seeding prescription is to populate halos (a) of mass $7.5 < \log (M_h/\Msun) < 8$, (b) in the redshift range $8 < z < 17$, (c) with IMBH in the mass range $4.75 < (\log M_\bullet/\Msun) < 6.25$. 

Although the present study combines the physics of SMS evolution and DCBH formation/growth with a well-tested cosmological scenario to derive the mass function of IMBH seed for the first time, it needs to be improved and complemented under many aspects.  

First, we have not attempted to constrain the formation efficiency of IMBH inside putative host halos. This would require the knowledge of the LW radiation field and a solid determination of $ J_{\nu,c}^\bullet$ during their formation epoch. Fortunately, given the very narrow mass range of the IMBH host halos ($7.5 < \log M_h < 8$)  the LW intensity can be factorized safely. 

Second, to follow the feedback regulated growth of DCBH we have assumed that the total mass of their hosting haloes remain constant during this short accretion phase. Using our merger tree model we found 
that the average time after which DCBH hosts experience a minor or major merger is respectively equal to $\approx 60$~Myr and $90$~Myr. Hence the approximation is good for our fiducial fertile minihalo case, 
as on average the accretion phase lasts for $\langle \Delta T_{accr}\rangle \approx 50$~Myr. However, 
this assumption may affect the IMBH IMF obtained for the sterile minihaloes as in this case $\Delta T_{accr}\approx 100$~Myr.

Third, a very interesting remaining question is the final fate of the population of IMBH that do not merge into SMBHs. As we have discussed, the IMBH seeds at formation are located inside dark matter halos that have lost all of their gas. Some of these systems will be able to re-accrete gas and turn it into stars (the raining gas is progressively more likely to be polluted); others might be included in larger halos and their IMBH merge with other black holes. Finally, some of them could remain isolated and dead, thus becoming virtually undetectable.   

As a last remark, we stress that during the feedback-regulated growth we have assumed spherically symmetric accretion. Although we have given arguments in support of this assumption, it is unclear if accretion might go through a disk that could become thermally unstable (e.g. because of \HH formation), form stars and SNe, thus stopping the IMBH growth. We plan to address these issues, which require dedicated high.resolution numerical simulation, in a forthcoming study. 

On the observational side, our scenario can have important implications. If a prolonged, obscured phase of DCBH growth exists, this might explain the puzzling near-infrared cosmic background fluctuation excess and its recently detected cross-correlation with the X-ray background \citep{Cappelluti13}, which might imply that an unknown faint population of high-$z$ black holes could exist (\citealt{Yue13a};\citealt{Yue13}). In addition, hints of a pervasive presence of IMBH in the center of nearby dwarf galaxies have been
convincingly collected by \cite{Reines13}. \cite{Rashkov14} also pointed out that about 70-2000 (depending on the assumptions made on their dynamics) relic IMBHs should be present in the Galactic bulge and halo. These objects might be indirectly traced by the clusters of tightly bound stars that should accompany them.
Thus, our results might be a solid starting point to make more detailed predictions on these and other related issues, including of course the puzzling presence of supermassive black hole in the first billion year after the Big Bang.  

\section*{Acknowledgments} 
AF acknowledges financial support from PRIN MIUR 2010-2011 project, prot. 2010LY5N2T. SS acknowledges support from Netherlands Organization for Scientific Research (NWO), VENI grant 639.041.233. DRGS thanks the German Science Foundation (DFG) for financial support via the Collaborative Research Center (CRC)~963 on ''Astrophysical Flow Instabilities and Turbulence'' (project A12) and via the Priority Program SPP~1573 ''Physics of the Interstellar Medium'' (grant SCHL 1964/1-1).
\bibliographystyle{apj}
\bibliography{ref}

\newpage 
\label{lastpage} 
\end{document}